\documentclass[a4paper,11pt]{article}
\pdfoutput=1
\usepackage{jheppub}
\usepackage{subfig}
\usepackage{amsmath,amssymb,bm,graphicx,epsf,colordvi}
\usepackage{slashed}
\def\bea#1\eea{\begin{align}#1\end{align}} 
\usepackage{comment}
\usepackage{diagbox}
\usepackage{mathtools}
\usepackage{lipsum}
\usepackage{soul}
\usepackage{enumitem}
\usepackage{braket}
\usepackage{bm}
\allowdisplaybreaks 
\usepackage{color}
\usepackage[dvipsnames]{xcolor}

\usepackage{mathrsfs}
\usepackage{appendix}
\usepackage{bm}
\usepackage{lipsum}

\newcommand{\bef}{\begin{figure}[hbt]\centering}
\newcommand{\eef}{\end{figure}}

\usepackage{mathtools}
\usepackage{booktabs}
\usepackage{graphicx}
\graphicspath{ {./figure/} }
\usepackage{caption}
\usepackage{lipsum}

\newcommand{\beq}{\begin{equation}}
\newcommand{\eeq}{\end{equation}}
\def\bea#1\eea{\begin{align}#1\end{align}}

\def \be  {\begin{equation}}
\def \ee  {\end{equation}}
\def \ba  {\begin{eqnarray}}
\def \ea  {\end{eqnarray}}

\newcommand{\rmd}{\mathrm{d}}

\allowdisplaybreaks

\usepackage{graphicx}
\usepackage{float}

\usepackage{multicol}
\usepackage[dvipsnames]{xcolor}

\def\Fig#1{Fig.~{\ref{#1}}}
\DeclareRobustCommand{\Sec}[1]{Sec.~\ref{#1}}
\DeclareRobustCommand{\App}[1]{App.~\ref{#1}}
\DeclareRobustCommand{\Eq}[1]{Eq.~(\ref{#1})}

\usepackage{stackengine}

\title{Visualizing How the Structure of Large-Radius Jets Shapes Their Wakes}

\author[a, b]{Arjun Srinivasan Kudinoor,}
\affiliation[a]{Center for Theoretical Physics, Massachusetts Institute of Technology, Cambridge, MA 02139}
\affiliation[b]{Department of Applied Mathematics and Theoretical Physics, University of Cambridge, Cambridge, CB3 0WA, UK}

\author[c]{Daniel Pablos,}
\affiliation[c]{Instituto Galego de Física de Altas Enerxías IGFAE,
Universidade de Santiago de Compostela, E-15782 Galicia-Spain}

\author[a]{Krishna Rajagopal}

\abstract{The ATLAS collaboration has introduced and implemented a strategy for selecting, and measuring the properties of, large-radius jets composed from skinny subjets in heavy ion collisions at the LHC. We show how measurements of these jets teach us about the resolution length $L_{\rm res}$ of quark-gluon plasma (QGP) and can teach us how jet substructure shapes the wakes that jets excite in the droplets of QGP through which they pass.
We begin by using Hybrid Model calculations to reproduce experimental measurements published by ATLAS in 2023 of $R_{AA}$ 
for large-radius jets in PbPb collisions, identified via first reconstructing skinny anti-$k_t$ $R=0.2$
subjets and then clustering $R=1$ jets using them as constituents. We investigate how $R_{AA}$
 for these large-radius jets depends on the angle between the two skinny subjets involved in the final reclustering step of the $R=1$
 jet. We show how these observables depend on $L_{\rm res}$ and demonstrate that the published ATLAS data rule out $L_{\rm res}=\infty$: these data are inconsistent with any picture in which an entire parton shower loses energy coherently as if it were a single colored object. 
Demonstrating that the QGP can resolve partons within a parton shower is central to the broader program of using jet quenching measurements to probe QGP.  

We make further use of this setup by analyzing the response of the medium to the passage of large-radius $R=2$
 jets containing two skinny $R=0.2$
 subjets, produced in gamma-jet events, identified as above. We introduce novel jet-shape observables that allow us to visualize the angular shape of the soft hadrons originating from the wakes that wide jets with two skinny subjets excite in the droplet of QGP, as a function of the angular separation between the subjets. We find that even when the two hard subjets are 0.8 to 1.0 radians apart, a single broad wake is produced. Only when the two subjets are even farther apart is the presence of two sub-wakes revealed. We show that the way in which jet structure shapes jet wakes can be visualized with similar clarity in experimental data by measuring the observables we have introduced using only those hadrons with low transverse momenta. These observables thus offer a new and distinctive way of seeing jet wakes in heavy ion collision data.
}

\preprint{MIT-CTP/5823}

\begin{document}

\maketitle


\section{Introduction}
Heavy-ion collisions at RHIC and the LHC have revealed that the high temperature phase of QCD, conventionally called quark-gluon plasma (QGP), is in fact a strongly coupled liquid~\cite{PHENIX:2004vcz,BRAHMS:2004adc,PHOBOS:2004zne,STAR:2005gfr,Gyulassy:2004zy}. 
The bulk properties of QGP and the dynamics of 
the expanding, cooling, droplets of QGP produced in heavy-ion collisions
have been 
elucidated via an interplay between experimental and theoretical investigations over the past two decades, with increasing clarity and precision.
(For reviews, see Refs.~\cite{Muller:2006ee,Casalderrey-Solana:2007knd,dEnterria:2009xfs,Wiedemann:2009sh,Majumder:2010qh,Jacak:2012dx,Muller:2012zq,Heinz:2013th,Mehtar-Tani:2013pia,Shuryak:2014zxa,Akiba:2015jwa,Romatschke:2017ejr,Connors:2017ptx,Busza:2018rrf,Nagle:2018nvi,Cao:2020wlm,Schenke:2021mxx,Cunqueiro:2021wls,Apolinario:2022vzg,Harris:2023tti,Grosse-Oetringhaus:2024bwr}.)
Studies of the bulk properties of QGP 
have led to 
a successful phenomenology based upon relativistic hydrodynamical simulations as well as 
numerous remarkable insights, all building upon the discovery that QGP behaves as a strongly coupled liquid
whose specific shear viscosity $\eta/s$ is close to the remarkably small value $1/(4\pi)$ that it takes on in any gauge theory with a holographic dual in the limit of infinitely strong coupling and infinitely many colors~\cite{Policastro:2001yc,Kovtun:2004de}.
(For a review, see Ref.~\cite{Casalderrey-Solana:2011dxg}.)



A particularly compelling way to access detailed information about the microscopic structure and dynamics of this new phase of matter is via the study of heavy-ion collision events in which high-energy jets are produced. 
These hard parton showers are produced at the earliest moments of the collision when the Lorentz-contracted incident nuclei first overlap. They then evolve and propagate within the droplet of QGP produced in the same collision, and are witnesses to the evolution in space and time of the QGP along their path.
The strong interactions between the high-energy colored partons in a parton shower and the colored excitations of the QGP result in modifications to the parton shower and, consequently, modifications to the energy and structure of the measured jet that are generically referred to as jet 
quenching~\cite{dEnterria:2009xfs,Wiedemann:2009sh,Majumder:2010qh,Jacak:2012dx,Muller:2012zq,Mehtar-Tani:2013pia,Connors:2017ptx,Busza:2018rrf,Cao:2020wlm,Cunqueiro:2021wls,Apolinario:2022vzg}. Measurements of the modification of the structure of a jet due to its passage through QGP can teach us about the properties and structure of the QGP, including for example its resolution length $L_{\rm res}$, the example that will be relevant here.  
$L_{\rm res}$, a temperature-dependent property of QGP, is the length scale such that if two jet partons are separated by less than $L_{\rm res}$ they behave as if they were a single colored object transferring energy and momentum to the QGP coherently, whereas if they are separated by more than $L_{\rm res}$ they interact independently with the QGP. Loosely speaking, the QGP can ``see'' two partons
in a parton shower as distinct if and only if they are separated by more than $L_{\rm res}$.
We shall show that a new class of jet quenching observables introduced and measured recently by the ATLAS collaboration~\cite{ATLAS:2023hso} is particularly well-suited to constraining the value of $L_{\rm res}$.

The resolution length $L_{\rm res}$ is a key element in a larger context that is central to describing jets in heavy ion collisions, namely the description of how the sequential fragmentation of an initially highly virtual parton into a parton shower proceeds in space and time.
When a parton shower evolves in vacuum, where and when each splitting occurs has no consequence; all that matters is how the momentum is shared.  In a heavy ion collision, though, the dynamics of showering occurs within a droplet of QGP that is expanding and cooling, meaning that the temperature of the QGP that a parton in the shower interacts with depends on both time and place.
By using a formation time argument, one can ascribe a spacetime picture to the standard 
momentum space parton shower~\cite{Casalderrey-Solana:2011fza} by assigning a duration to each rung in the decay chain given by $2E/Q^2$, with $E$ and $Q$ being the energy and virtuality of that parton in the shower.
This approach underlies the Hybrid Model for jet quenching~\cite{Casalderrey-Solana:2014bpa,Casalderrey-Solana:2015vaa,Casalderrey-Solana:2016jvj,Hulcher:2017cpt,Casalderrey-Solana:2018wrw,Casalderrey-Solana:2019ubu,Hulcher:2022kmn}, in which it is used to keep track of production points and trajectories of each of the many partons in the shower. 
This spacetime picture of the parton shower is then embedded within
a space- and time-dependent solution to the hydrodynamics equations that describe the expanding cooling droplet of QGP, which in turn allows for the determination of the rate at which each parton in the shower loses energy and momentum to the strongly coupled QGP at each point in space and time, with its local temperature and fluid velocity.
In the Hybrid Model with $L_{\rm res}=0$~\cite{Casalderrey-Solana:2014bpa,Casalderrey-Solana:2015vaa,Casalderrey-Solana:2016jvj}, the splitting and spacetime evolution of the parton shower is determined entirely as if the shower were evolving in vacuum -- and it is this spacetime picture that gets modified by considering a nonzero QGP resolution length $L_{\rm res}$, as this serves 
to delay the energy-loss consequences of each splitting while the two partons formed at the splitting interact with the QGP as if they were a single color charge until they have separated by more than $L_{\rm res}$~\cite{Hulcher:2017cpt,Casalderrey-Solana:2018wrw,Casalderrey-Solana:2019ubu}.
This finite resolution effect is understood in perturbative QCD (pQCD) in the multiple soft scattering regime in terms of a critical coherence angle, $\theta_c$~\cite{Casalderrey-Solana:2011ule,Mehtar-Tani:2011hma,Armesto:2011ir,Mehtar-Tani:2012mfa,Casalderrey-Solana:2012evi,Dominguez:2019ges,Abreu:2024wka}, above which two charges of a color dipole act as independent sources of gluon radiation.  In the case where the interactions between the partons in the parton shower and the QGP in which they find themselves occurs as in a strongly coupled gauge theory plasma, as in the Hybrid Model that we shall employ throughout the present paper, a nonzero resolution length $L_{\rm res}$ can be understood 
as stemming from the physics of screening in a strongly coupled thermal plasma~\cite{Bak:2007fk}, with a resolution length of the order of the Debye screening length $\lambda_D$ which is $\sim 0.3/(\pi T)$ in the strongly coupled plasma of ${\cal N}=4$ SYM theory~\cite{Bak:2007fk} and $\sim (1-2)/(\pi T)$ in strongly coupled QGP~\cite{Hulcher:2017cpt}.
With this as context, we can see the importance of using experimental measurements to constrain the value of $L_{\rm res}$: this property of QGP governs how the spacetime structure of a parton shower is modified if that parton shower evolves within a droplet of QGP rather than in vacuum.

We have focused so far on asking how parton showers, and the resulting jets, are modified via their passage through a droplet of QGP.
It is just as important to ask how the droplet of QGP is modified via the passage of a jet through it. The passing jet loses momentum and energy 
to the QGP, exciting a wake in the expanding cooling droplet. The characterization of these
perturbations in experiments 
gives us access to the dynamics of QGP. The jet pushes a region of the droplet of QGP somewhat out of equilibrium, and seeing how the resulting disturbance evolves gives us access to information about the excitations and dynamics of QGP that is complementary to what can be learned by asking how the parton shower is modified.
The second 
main objective of the analysis presented in this paper is to provide new ways with which to visualize and analyze these fluid perturbations, referred to as jet wakes, using novel jet substructure techniques built upon the same new observables introduced and measured by ATLAS~\cite{ATLAS:2023hso}.

Energetic jets as reconstructed from experimental data are objects whose 
structure encodes physics at many scales, ranging from the very high momentum transfer in the hard scattering from which the jets originate through the intermediate scales associated with the DGLAP evolution of the parton shower to the lower scales associated with the exchange of momentum between partons in the shower and the strongly coupled QGP to the lowest scales associated with the temperature and flow velocity of the droplet of QGP itself and the jet wakes excited in it.
Note that what an experimentalist reconstructs and calls a jet must include hadrons originate from the freezeout of the wake in the droplet of QGP as well as hadrons originating from the fragmentation of the parton shower because the wake carries momentum in the jet direction, the momentum lost by the parton shower.
We shall show that the observables introduced by ATLAS~\cite{ATLAS:2023hso} that realize a strategy for finding large-radius jets composed of skinny subjets give us new access to information about the QGP resolution length $L_{\rm res}$ via suitably differential measurements of how such jets are quenched and at the same time offer new ways to visualize jet wakes and to see how their shapes are governed by the separation between the skinny subjets .



Before proceeding, it is worth noting that in modern jet quenching models the response of the droplet of QGP to the passage of a parton shower through it is described using one of two distinct frameworks. The direction that we shall not pursue is based upon accounting for the partons from the medium that recoil after having been kicked via an elastic scattering process with a parton from the jet 
shower~\cite{DEramo:2012uzl,Zapp:2013vla,He:2015pra,Cao:2016gvr,He:2018xjv,DEramo:2018eoy,Park:2018acg,Dai:2020rlu,Ke:2020clc,JETSCAPE:2021ehl,Hulcher:2022kmn,Luo:2023nsi,Pablos:2024muu}. 
To date, these approaches to
modeling back-reaction in this way assume that the process can be described perturbatively, which is something that is only guaranteed when the momentum transfer is high enough. 
In most (but not all) cases they also neglect the possibility of subsequent interactions between the recoiling partons and the medium, which are in fact likely to be significant in a strongly coupled liquid. 
The second approach, that we shall employ, is based upon assuming that
the momentum lost by the parton shower (for example via the aforementioned recoils) quickly becomes hydrodynamic excitations of the droplet of QGP (for example via the strong interactions between recoiling medium partons and the strongly coupled fluid whence they originate)
and hence
studying the response of the fluid QGP energy-momentum tensor to the injection of energy and momentum due to its interactions with the energetic partons in the parton shower~\cite{Casalderrey-Solana:2004fdk,Ruppert:2005uz,Renk:2005si,Casalderrey-Solana:2006lmc,Betz:2008ka,Neufeld:2008fi,Chesler:2007an,Gubser:2007ga,Gubser:2007ni,Gubser:2007xz,Chesler:2007sv,Li:2010ts,Tachibana:2014lja,Tachibana:2015qxa,Casalderrey-Solana:2016jvj,Yan:2017rku,Okai:2017ofp,Chen:2017zte,Tachibana:2017syd,Chang:2019sae,Casalderrey-Solana:2020rsj,Chen:2020tbl,Pablos:2022piv,Yang:2021qtl,Yang:2022nei,Yang:2025dqu}. This is the more appropriate framework when momentum transfers are in the non-perturbative regime and when we take advantage of the fact that hydrodynamics is used as the effective long-wavelength description of the system. In the present work, we will only focus on the phenomenology of the hadrons that originate from the freezeout of jet-induced hydrodynamic wakes, leaving the inclusion of elastic recoils and the possibility that they may modify the shapes of the jet wakes to future work.

In this paper, we shall use Hybrid Model calculations to show how to use experimental measurements of a new class of jet substructure observables pioneered by the ATLAS collaboration~\cite{ATLAS:2023hso} 
to constrain the value of $L_{\rm res}$ and to 
visualize and study how the substructure of the jets shapes their wakes.
The ATLAS approach
consists of first identifying skinny jets (to be referred to as skinny subjets) reconstructed using the anti-$k_t$ algorithm with a small value of $R$ 
and then using these objects as the constituents from which to reconstruct a single large-radius jet.
This yields a selection strategy for identifying an ensemble of (very) large-radius jets that each contain a specified number of skinny subjets that is realizable in the analysis of experimental data.
The key handle that this observable offers is the possibility to control the angular separation between the skinny subjets within the large-radius jet. 
After first describing the  Hybrid strong/weak coupling Model that we shall employ throughout to simulate jet quenching physics in Section~\ref{sec:model} and providing details about the simulations, we describe the observables that we shall calculate and the corresponding simulated data sets in Section~\ref{sec:setup}.
In Section~\ref{sec:suppression},
we investigate how large-radius jets with different
numbers of skinny subjets and with varying separation between the skinny subjets
are suppressed in heavy-ion collisions as compared to proton-proton collisions. 
We do so employing three different values of the QGP resolution length: $L_{\rm res}=0$, $2/(\pi T)$, and $\infty$, 
and compare our predictions to published ATLAS measurements~\cite{ATLAS:2023hso}.
We show that the fact that in the ATLAS data large radius jets containing multiple skinny subjets are far more suppressed than those containing only a single skinny subjet is in strong disagreement with the results of our calculations with $L_{\rm res}=\infty$. The disagreement is sufficiently stark that we expect that the ATLAS data are inconsistent with any approach that (as in the Hybrid Model with $L_{\rm res}=\infty$) assumes that entire parton showers lose energy and momentum cohrently as if they were single energetic colored objects.
We comment
on the discriminating power that such measurements have in determining the value of $L_{\rm res}$, observing that today's data mildly favor $L_{\rm res}=2/(\pi T)$ over $L_{\rm res}=0$ and discussing how measurements to come including those with smaller values of the separation between skinny subjets
can be optimized to further constrain the value of $L_{\rm res}$.

In Section~\ref{sec:substructure},
we perform a completely novel study with the aim of 
visualizing and analyzing the consequence, the result, the measurable imprints, of the suppression of large radius jets 
containing two skinny subjets as a function of the angle between the two subjets.
How do the wakes deposited in a droplet of QGP by these large radius jets with differing substructure --- with differing separation between their two skinny subjets ---
vary?  It is well known that after freezeout jet wakes become a broad cloud of soft hadrons centered on the jet axis, but how does this statement get modified when the jet in question has two well-separated skinny subjets?
We use the Hybrid Model to provide answers to questions like this, and in so doing open new pathways via which experimentalists can obtain distinctive visualizations
of jet wakes in data, and can investigate how jet substructure shapes jet wakes.
Given the relatively broad angular distribution of the soft hadrons originating from a jet wake, 
pursuing these questions would be nearly impossible using conventional, moderate-$R$, jets.  The strategy pioneered by ATLAS for selecting exactly those large-radius jets that are composed from skinny subjets makes this investigation possible. 

We introduce a new class of differential jet shape observables, suitable for the analysis of large radius jets containing two well-separated skinny subjets, and use the Hybrid Model (in particular the ability to turn jet wakes off and on which of course is impossible to do in experimental data) to show that the measurement of these observables using hadrons in a range of soft momenta like $0.7<p_T<1.0$~GeV yields a good proxy for visualizing the shapes of the wakes of these jets.
We find that even when the skinny subjets are separated by as much as 1.0 radians, the soft hadrons from the wakes of the two subjets overlap, forming a conventional single-peaked distribution centered on the axis of the large-radius jet.  Only when the skinny subjets are separated by even larger angles do we see separated 
wakes, namely a double-peaked distribution of soft hadrons centered separately on each skinny subjet.


Despite many works over the past decade highlighting the importance of medium response in describing jet quenching observables including many substructure observables, it is only very recently that unambiguous evidence of this phenomenon has been found in experiments. 
Almost all of the momentum carried by the wake of a jet (and lost from the parton shower) is carried by a region of fluid behind the jet moving in the direction of the jet.  Comparing the hadronization of a droplet of QGP including a jet wake to that of a droplet of QGP with no jet, the wake corresponds to an excess of soft hadrons in a broad range of angles around the jet as well as a depletion in the number of soft hadrons in the opposite direction in the transverse plane.
The first measurements seeking evidence for this depletion on the photon-side in photon-jet events were made by ATLAS~\cite{ATLAS:2024prm} and very recently 
evidence for this effect has been presented by CMS~\cite{CMS:2024fli} in $Z$-hadron correlation measurements without jet reconstruction, finding a depletion around the $Z$.
CMS has also shown
that models without medium response (for example the Hybrid Model with jet wakes artificially turned off) 
fail to describe the data while models that include the effects on the soft hadron distribution coming from jet wakes see phenomena that are comparable to that seen in the data.
%
%
This important finding adds confidence to the quantitative investigation of the contribution of jet wakes to many other observables in addition.

The new differential jet shape observables that we have introduced in this paper, employed together with the selection of large radius jets containing two well-separated skinny subjets using the methods pioneered by ATLAS,
point the way to additional, distinctive, visualizations of the distribution of soft hadrons originating from jet wakes.  If (we hope when) experimentalists see a single overlapping soft wake enveloping two well-separated skinny subjets in a sample of large radius jets in photon-triggered events fission into distinct soft wakes around the subjets only when their separation grows beyond one radian in angle, this will be a striking and distinctive measurement.
Beyond such a first measurement, when the tools and methods that we have introduced in this work are employed in future high-precision comparisons between (further improved) theoretical predictions and (higher statistics) experimental measurements in which the orientation of the skinny subjets relative to the event-plane of the collision as well as their angular separation can both be controlled, we can look forward to learning about the interference between overlapping subjet wakes and about the interplay between jet wakes and the background radial flow of the expanding droplet of QGP.



\section{The Hybrid Strong/Weak Coupling Model}
\label{sec:model}
\subsection{A Brief Description of the Model} \label{sec:hybridintro}
The hybrid strong/weak coupling model for jet quenching, or simply the Hybrid Model, described in Refs.~\cite{Casalderrey-Solana:2014bpa,Casalderrey-Solana:2015vaa,Casalderrey-Solana:2016jvj,Hulcher:2017cpt,Casalderrey-Solana:2018wrw,Casalderrey-Solana:2019ubu,Hulcher:2022kmn}, is a theoretical framework that aims to describe the multi-scale processes of jet formation and evolution within the QGP. The production and showering of energetic partons from which jets originate occur at scales at which QCD is weakly coupled. However, the most common interactions amongst constituents of the QGP and interactions between partons in the jet shower and medium involve momentum scales that are of the order of the QGP temperature $T \sim \Lambda_{\rm QCD}$. At these scales, QCD is strongly coupled and the QGP behaves as a strongly coupled liquid. The essence of the Hybrid Model lies in describing these different regimes using the methods that are appropriate given the dominant energy scales involved. 

In a heavy-ion collision, jet formation occurs at time scales that are far before the formation of the QGP, since the hard scattering process from which a high energy jet originates is characterized by a high virtuality scale $Q \sim p_T^{\rm jet} \gg T$. The high virtuality partons from such hard scattering processes relax by successive splittings that result in  jet showers. Due to the high virtuality scale that characterizes the initial hard scattering and subsequent shower, the Hybrid Model describes these processes using perturbative QCD. The splittings that result in the jet shower are determined by the perturbative, high-$Q^2$, DGLAP evolution equations. The Hybrid Model employs PYTHIA 8~\cite{Sjostrand:2014zea}, with initial state radiation but not multi-parton interactions, to describe the perturbative evolution of energetic partons in a jet shower. Modified parton distribution functions following the EPS09 parametrization~\cite{Eskola:2009uj} are used to account for initial state nuclear effects that are responsible for the different production rates of hard processes in PbPb collisions with respect to proton-proton collisions. Each parton in the shower is assigned a lifetime which is defined to be the time between the moment a parton is created at a splitting and the moment that it itself splits. This lifetime is given by the expression ~\cite{Casalderrey-Solana:2011fza}
\begin{equation} \label{eq:lifetime}
    \tau = 2E/Q^2,
\end{equation}
where $E$ is the parton's energy and $Q$ is its virtuality. Assigning a lifetime to each parton in the hard parton shower in this way allows one to endow each parton shower generated by PYTHIA with a spacetime structure. These showers are then embedded into a space- and time-dependent background that describes an expanding and cooling droplet of QGP obtained via the relativistic viscous hydrodynamics calculations of Ref.~\cite{Shen:2014vra}. The origin of the hard event in the plane transverse to the direction of the collision is chosen probabilistically using the Glauber model, according to the density distribution of the nuclear overlap for the same centrality class, and its azimuthal direction is 
chosen randomly. The hydrodynamic QGP background then provides local QGP temperatures and fluid velocities at each spacetime point, necessary for computing the modifications to parton showers that are imparted by their interaction with the expanding and cooling QGP droplet.

Although jet production and fragmentation are weakly coupled QCD processes, the physics of the QGP and the dominant interactions between hard partons and the medium are not. 
So, the Hybrid Model models these interactions by assuming that an energetic quark traversing the QGP droplet loses energy at a rate equivalent to that of a light quark traversing strongly coupled plasma in ${\cal N}=4$ supersymmetric Yang--Mills theory in the limit of infinite coupling and large $N_c$. For an energetic color charge in the fundamental representation 
that has traveled a distance $x$ in the fluid rest frame, this energy loss rate, calculated holographically in Refs.~\cite{Chesler:2014jva,Chesler:2015nqz},  takes the form
\begin{equation}
\label{eq:elossrate}
   \left. \frac{dE}{dx}\right|_{\rm strongly~coupled}= - \frac{4}{\pi}\, \frac{E_{\rm in}}{x_{\rm stop}}\, \frac{x^2}{x_{\rm stop}^2} \frac{1}{\sqrt{1-(x/x_{\rm stop})^2}} \quad ,
\end{equation}
where $E_{\rm in}$ is the initial energy of the energetic color charge before it loses any energy to the strongly coupled plasma and $x_{\rm stop}\equiv  E_{\rm in}^{1/3}/(2{T}^{4/3}\kappa_{\rm sc})$ is the maximum distance the parton can travel within the plasma before completely thermalizing. $\kappa_{\rm sc}$ is a dimensionless free parameter 
that quantifies the degree of strong coupling between energetic partons and the medium. Formally, it depends on the t'Hooft coupling $\lambda = g^2 N_c$, the details of the strongly coupled gauge theory, and the initial conditions under which the energetic parton originated. In ${\cal N}=4$ supersymmetric Yang-Mills theory, $\kappa_{\rm sc}$ can be calculated and is proportional to $\lambda^{1/6}$. In the Hybrid Model, where the goal is to describe the propagation of energetic partons through the strongly coupled QGP described by QCD as produced in heavy ion collisions, $\kappa_{\rm sc}$ is chosen by
fitting to experimental measurements of high-$p_T$ hadron and jet suppression from the LHC~\cite{Casalderrey-Solana:2018wrw}. 
The fit results in a thermalization distance $x_{\rm stop}$ that is longer by a factor of 3-4 for QGP in QCD when compared to that for strongly coupled plasma in ${\cal N}=4$ supersymmetric Yang--Mills theory at the same 
temperature~\cite{Casalderrey-Solana:2014bpa,Casalderrey-Solana:2015vaa,Casalderrey-Solana:2016jvj,Casalderrey-Solana:2018wrw}. Additionally, note that the holographic stopping distance of a gluon is reduced by a factor of $(C_A/C_F)^{1/3} = (9/4)^{1/3}$ when compared to that of a quark~\cite{Gubser:2008as}, where $C_A$ and $C_F$ are the Casimirs of the adjoint and fundamental representations of the color gauge group, respectively. This means that
\begin{equation} \label{eq:kappagluon}
   \kappa_{\rm gluon} = (9/4)^{1/3} \kappa_{\rm sc}.
\end{equation}
In the Hybrid Model, we take a space- and time-dependent temperature $T$ from the hydrodynamic background and apply Eq.~(\ref{eq:elossrate}) to each parton in the jet shower at every timestep, where $x_{\rm stop}$ is evaluated using the temperature $T$ at each parton's spacetime location.

Having examined how energetic partons lose energy as they traverse the QGP in the Hybrid Model, one may ask where this lost energy goes; energy and momentum must be conserved, after all. This lost momentum and energy is deposited into the QGP droplet, inducing perturbations of the stress-energy tensor of the strongly coupled liquid QGP. The injected energy hydrodynamizes at distance scales on the order of $1/T$ (at strong coupling), which induces a wake in the plasma that carries the lost momentum and energy of the jet shower. This wake evolves hydrodynamically as the droplet of QGP expands, flows, and cools. When the QGP, including the wake(s) contained within it, reaches the freeze-out hypersurface, it hadronizes into thousands of soft hadrons. A subset of these hadrons are the result of the jet wake(s) hadronizing at freezeout; these hadrons carry a net momentum in the jet direction, corresponding to the momentum lost to the fluid by the jet partons. What an experimentalist later reconstructs as a jet includes hadrons originating from the hadronization of the (degraded) parton shower and hadrons originating from the freezeout of the wake in the droplet of QGP.

In the Hybrid Model, the spectrum of the hadrons from jet-induced wakes is obtained by employing the Cooper--Frye prescription~\cite{PhysRevD.10.186} and using a set of simplifying assumptions that enable a simple analytic estimate of this contribution to the final state hadron spectra~\cite{Casalderrey-Solana:2016jvj}. First, the background on which the wake propagates is assumed to be longitudinally boost invariant. Second, the perturbation is assumed to stay close in rapidity to the rapidity at which it was deposited, namely the rapidity of the jet. Third, the 
wake is assumed to be a small perturbation to the hydrodynamic evolution of the droplet of QGP prior to freezeout, which is a good approximation. 
Fourth, the perturbation to the phase-space distribution of hadrons immediately after freezeout coming from the temperature and flow perturbations at freezeout caused by the wake is assumed to be small for all hadron momenta.
This is a good approximation for the softest hadrons, but 
need not be a good approximation for hadrons with momenta of order a few GeV, where the unperturbed Cooper-Frye spectra are sufficiently Boltzmann suppressed that the contribution due to hadrons originating from jet wakes need not be a small perturbation.
For a jet with azimuthal angle $\phi_j$ and rapidity $y_j$ that has lost transverse momentum and energy $\Delta p_{T}$ and $\Delta E$ to the plasma, upon making all of these approximations the momentum spectrum of the hadrons at freezeout resulting from the jet-induced wake is given by the closed-form expression~\cite{Casalderrey-Solana:2016jvj}
\begin{equation}
\label{eq:onebody}
\begin{split}
E\frac{\rmd\Delta N}{\rmd^3p}=&\frac{1}{32 \pi} \, \frac{m_T}{T^5} \, \textrm{cosh}(y-y_j)  e^{-\frac{m_T}{T}\, \textrm{cosh}(y-y_j)} \\
 &\times \Bigg\{ p_{T} \Delta p_{T} \cos (\phi-\phi_j) +\frac{1}{3}m_T \, \Delta M_T \, \textrm{cosh}(y-y_j) \Bigg\} \, ,
\end{split}
\end{equation}
where $y$, $\phi$, $p_{T}$, and $m_T$ are the rapidity, azimuthal angle, transverse momentum, and transverse mass of the emitted wake-hadrons and where $\Delta M_T\equiv \Delta E/\cosh y_j$. In order to generate the wake hadrons, Eq.~(\ref{eq:onebody}) is repeatedly sampled until the sum of momenta of all these particles matches the four-momentum deposited by the corresponding wake-generating parton.
Already from the first comparisons between the predictions of Eq.~(\ref{eq:onebody})
to then available experimental data, it has been clear that this simple analytical expression underestimates the production of semi-hard hadrons, say those with 2 GeV $< p_T <$ 4 GeV, originating from jet wakes and, correspondingly by momentum conservation, overpredicts the number of such hadrons with the lowest momenta~\cite{Casalderrey-Solana:2016jvj}.

Inspection of Eq~(\ref{eq:onebody})  reveals that the spectrum can become negative in an azimuthal region $\phi$ that is opposite to the azimuthal angle $\phi_j$ of the jet. 
This can be understood as follows.
Before freeze-out, the energy and momentum in the hydrodynamic wake is carried by sound modes and diffusive
modes.  The diffusive modes describe the moving fluid behind the jet, fluid that is moving in the direction of the jet.  Almost all of the momentum lost by a jet is in fact carried by this moving fluid, hence by the diffusive modes.  The sound waves, which describe compression and rarefaction of the fluid and the corresponding perturbations to its temperature carry little momentum.
At freezeout, the boosted fluid cells corresponding to the diffusive modes, to the moving fluid behind the jet,
are translated by the Cooper--Frye prescription into
an enhancement of soft hadrons in the direction of the jet and a depletion of soft hadrons in the direction opposite the jet in the transverse plane, when compared to an unperturbed background. 
This is what Eq~(\ref{eq:onebody}) describes.
In the Hybrid Model, these features of
Eq~(\ref{eq:onebody}) are implemented by adding hadrons (sometimes referred to as ``positive hadrons'') in the regions of phase-space where the distribution in Eq.(\ref{eq:onebody}) is positive, and by adding ``negative hadrons'' hadrons where Eq~(\ref{eq:onebody}) is negative. 
When calculating any observable that is linear in the energies of particles within a calorimetric cell, like jet-$p_T$ or the jet shape, one must simply add the energies of particles in the cell, while treating the energies of the negative hadrons to be negative.
``Negative hadrons'' are to be understood as the absence of a particle with that momentum which would be present if not for the wake. In essence, by boosting some QGP that it has passed through into motion in its own direction the jet reduces the spectrum of hadrons going opposite to its own direction coming from the freezeout of the droplet of QGP at the same time that it increases the spectrum of hadrons from the QGP in its own direction.
In the Hybrid Model it is possible to artificially turn off or turn on the contribution coming from jet wakes, and in fact it is possible (even more artificially) to separately turn off or turn on the
positive and negative contributions from jet wakes, that we can refer to in brief as the positive and negative wakes.

As parton showers traverse the QGP droplet and lose energy (that is then deposited into the jet wakes), some of the partons in these showers lose all their energy to the plasma. The 
energy and momentum that (in PYTHIA) 
would have been carried by such a parton are described entirely by 
the freezeout of the wakes they have created, using Eq~(\ref{eq:onebody}). 
At freezeout, though, many of the (more energetic) partons
in the shower remain, as they have not lost all of their energy.
These shower partons 
are hadronized via the Lund string model implemented in PYTHIA 8. Since the energy and momentum lost by partons to the QGP is deposited into the wake distribution as described by Eq.~(\ref{eq:onebody}), energy and momentum are conserved in each event. 

In the Hybrid Model, we do not consider any additional modifications to the hadrons originating from the degraded parton shower or to the hadrons originating from the wake after freeze-out.

In the Hybrid Model, we are able to distinguish between hadrons that result from a jet wake and hadrons that result from jet fragmentation (i.e.~from the degraded parton shower). However, there is no way to make this distinction among the hadrons that are used to reconstruct jets in experimental data; both hadrons from the wake and hadrons from jet fragmentation will be included in reconstructed jets. The ability to make this distinction in the Hybrid Model enables us to isolate and study the behavior of jet wakes as prescribed by the model. Our model analysis can then illuminate avenues for identifying and studying jet wake observables in experiment. That is the spirit of this study.

\subsection{Resolution Length of the QGP} \label{sec:lres}
There are multiple additional physical effects that previous studies have included in the Hybrid Model. These include the soft transverse kicks (with a Gaussian distribution for the transverse momentum transfer) experienced by jet partons traversing strongly coupled plasma~\cite{Casalderrey-Solana:2016jvj}; semi-hard elastic scatterings off medium constituents, assumed perturbative, known as Moli\`ere scatterings~\cite{Hulcher:2022kmn}; and the possible inability of the medium to resolve different partons in a jet shower as distinct sources of energy loss only if such partons are separated by less than a characteristic length, called the QGP resolution length~\cite{Hulcher:2017cpt}. The former two of these effects have not been included in this study, whereas the effect of the QGP resolution length is included and discussed.

As was first presented in Ref.~\cite{Hulcher:2017cpt}, the resolution length of QGP must be comparable to or possibly smaller than the Debye screening  length $\lambda_D$ for color charges in the medium, which for a strongly coupled plasma with temperature $T$ is $\lambda_D \sim 1/T$. In the specific case of a plasma in ${\cal N} = 4$ supersymmetric Yang--Mills theory, $\lambda_D \approx 0.3/(\pi T)$~\cite{Bak_2007}. Since a QCD plasma has fewer degrees of freedom, we can expect $\lambda_D$ to be larger by some factor, in the regime of temperatures where QCD is strongly coupled. In this study, following Ref.~\cite{Hulcher:2017cpt} we examine a case where $L_{\rm res} = 2/(\pi T)$, in addition to cases 
where $L_{\rm res} = 0$ and $L_{\rm res} = \infty$.
$L_{\rm res}=\infty$ means that the QGP cannot resolve any substructure within a jet at all; it ``sees'' the entire jet as if it were a single parton carrying all the momentum of the jet.
$L_{\rm res}=0$ means that the QGP ``sees'' every parton in the shower as a distinct source of energy and momentum; this is the default assumption in the Hybrid Model when we do not turn on the possibility of a nonzero QGP resolution length.

In practice, the resolution length $L_{\rm res}$ characterizes a length scale such that if two partons that result from the same splitting are separated by a length smaller than $L_{\rm res}$, then they will lose energy to the plasma -- and produce a wake -- as if they were a single parton. For example, suppose an arbitrary parton in the jet shower undergoes a 1 to 2 splitting after propagating for a time $\tau_0$. Suppose the two daughter partons take time $\tau_{\rm sep}$ to separate by a length greater than the QGP resolution length. Since the medium cannot resolve the two daughter partons as individual charges until they are separated by at least $L_{\rm res}$, the parent particle is quenched according to Eq.~(\ref{eq:elossrate}), with its summed color charge, 4-momentum, and decreasing total energy, as if it had propagated for a time $\tau_0 + \tau_{\rm sep}$. (Effectively, this amounts to a modified lifetime of the parent parton, i.e. the time from when the parent parton begins to lose energy independently to when its daughter partons are first resolved by the medium.) In essence, accounting for the QGP resolution length modifies 
the spacetime structure of a parton shower in the Hybrid Model.



Ref.~\cite{Hulcher:2017cpt} details the implementation of the QGP resolution length parameter in the Hybrid Model, as well as its effects on jet shapes and jet fragmentation functions. It was shown that implementing a nonzero resolution length in the Hybrid Model increased the likelihood that particles carrying a small fraction of the jet energy at larger angles from the jet axis would survive their passage through the QGP. In our study, we examine the effects of zero, finite nonzero, and infinite resolution lengths on a combination of jet suppression and jet substructure observables. 


\section{Simulation Setup} \label{sec:setup}
\subsection{Simulation Details}
The studies in this paper are performed using a variety of samples generated using the hybrid strong/weak coupling model framework as described in Section~\ref{sec:hybridintro}. These samples include proton-proton collision events (in vacuum, without the presence of the QGP medium) and PbPb collision events (with the presence of the QGP medium); inclusive jet events (with selection criteria defined by the momenta and rapidity of a jet in the event)
and events with jets produced in association with a hard photon ($\gamma$-tagged jets, with selection criteria defined by the momenta and rapidity of the hard photon).
In all cases, we have generated samples of collision events whose QGP resolution lengths are 0, $2\pi/T$, and $\infty$, where $T$ is the temperature of the plasma.
The value of $\kappa_{\rm sc}$ for $L_{\rm res}=0$ and $L_{\rm res}=2/(\pi T)$ was set to be 0.404 and 0.438, respectively. These values are taken from the global fit in Ref.~\cite{Casalderrey-Solana:2018wrw}, performed using then available experimental data on inclusive jet and single inclusive hadron suppression in central collisions at RHIC and LHC. The corresponding value for $L_{\rm res}=\infty$ has been set to $\kappa_{\rm sc}=0.5$ by imposing that jet suppression for $R=0.4$ anti-$k_T$ jets coincides with the other choices of the value of the QGP resolution length at around $p_T\sim 100$ GeV in central PbPb collisions~\cite{Casalderrey-Solana:2019ubu}. These values of $\kappa_{\rm sc}$ correspond to the scenario where the value of the pseudocritical temperature below which there is no energy loss is chosen to be $T_c=145$ MeV.
Each inclusive jet sample contains $\sim$ 1 million events and each $\gamma$-jet sample contains $\sim$ 1.5-2 million events, all for collisions with a center of mass energy of $\sqrt{s} = 5.02$ TeV per nucleon-nucleon collision. The heavy-ion events simulated correspond to the 0--10\% most central collisions in Section~\ref{sec:suppression} and to the 0--5\% centrality class in Section~\ref{sec:substructure}.

In the inclusive jet samples, we set the minimum $p_{T}$ of the initial hard scattering from which the jet showers originate to be $\widehat{p}_{T, \rm min}=50$ GeV. In the $\gamma$-jet samples, with $\widehat{p}_{T, \rm min}=100$ GeV, the minimum $p_{T}$ for the trigger photon used to select the events is 150 GeV. Both photons from the hard scattering and those produced during the parton shower evolution may be selected as event triggers. The PYTHIA version used is 8.244, with NNPDF2.3 parton distribution functions (PDFs) for proton-proton collisions. For PbPb collisions, the PDFs are modified according to the EPS09LO~\cite{Eskola:2009uj} nuclear PDFs. While initial state radiation is turned on, multi-parton interactions are turned off. The default value of \texttt{TimeShower:pTmin} has been modified to 1 GeV.


\subsection{Jet Reconstruction} \label{sec:reconstruction}
In this study, there are two types of full-hadron level jets that were reconstructed using \texttt{FastJet 3.4.1}~\cite{Cacciari:2011ma}. The first type of jets were reconstructed (in a standard fashion) by clustering final state hadrons -- generated using PYTHIA and the Hybrid Model framework described above -- using the anti-$k_{T}$ algorithm~\cite{Cacciari:2008gp} with radius parameter $R$ and an acceptance of $|y| < 2.0$. The second type of jets in this study are large-radius jets reconstructed via a procedure introduced by the ATLAS collaboration that is described in full detail in Ref.~\cite{ATLAS:2023hso}
These large-radius jets are reconstructed by first reconstructing small-radius anti-$k_t$ $R = 0.2$ jets satisfying $|\eta| < 3.0$ and $p_T > 35$ GeV that we shall often refer to as ``skinny subjets'' and then clustering these objects into large-radius jets with $|y|<2.0$ using the anti-$k_t$ algorithm with a radius parameter $R$ that can be much larger than 0.2. ATLAS has used this procedure to reconstruct $R=1.0$ jets from the skinny subjets; we have extended their procedure to values of $R$ as large as 2.0.
We call the small-radius anti-$k_t$ $R = 0.2$ jets within each large-radius jet a ``skinny subjet" of the corresponding large-radius jet.  Unless otherwise specified, no cuts on the $p_{T}$ of constituent hadrons that enter any jets were applied. Since the $k_t$ algorithm~\cite{Catani:1993hr,Ellis:1993tq} tends to combine the hardest constituents of a jet last, we use it to re-cluster the constituents of each large-radius jet so that we may study observables that depend on the hardest splitting of anti-$k_t$ subjets within each large-radius jet.

After reproducing the ATLAS analyses
of large-radius jets in the Hybrid Model with different values of $L_{\rm res}$ in \Sec{sec:suppression},
in \Sec{sec:substructure} we study the substructure of such large-radius jets in $\gamma$-jet events. Such events are characterized by the presence of an isolated prompt photon that is roughly back-to-back, at leading order, with a jet(s). At leading order in perturbative QCD, prompt photons are produced in $2 \rightarrow 2$ scatterings, with minimal surrounding hadronic activity. In contrast, fragmentation photons are produced during jet fragmentation and will be surrounded by many hadrons resulting from the jet shower. In order to identify a $\gamma$-jet event, i.e. an event with an isolated and energetic prompt photon, a number of cuts were applied. First, the event must contain a high-energy photon with $p_T > 150$ GeV and $|\eta^\gamma| < 1.44$ that is also ``isolated" which, we define as having less than 5 GeV of transverse energy in a cone of $R = 0.4$ around the photon. Amongst all such possible photons in an event, we only select the photon with the highest transverse momentum. Since we aim to select jets that recoil from this photon, we only select those anti-$k_t$ $ = 0.2$ subjets with $\Delta \phi > 2\pi /3$ in relation to the highest-$p_T$ photon. Finally, the selected small-radius anti-$k_t$ $R = 0.2$ subjets with $p_T > 35$ GeV and $|\eta| < 3.0$ are clustered into large-radius jets.

In \Sec{sec:substructure} we shall define a new class of differential jet shape observables with which to characterize 
large-radius jets with $R=2.0$ made up of exactly two skinny subjets. For these shape observables we shall include all hadrons within a radius 2 of the large-radius jet axis, whether or not each hadron is a part of one of the two skinny subjets.



\section{Large-Radius Jet Suppression} \label{sec:suppression}
In this Section, we study and compare calculations of jet suppression in inclusive jet events, as a function of their $p_T$, their substructure properties, and the role played by the QGP resolution length. We first present $R_{\rm AA}$ calculations for standard anti-$k_t$ $R = 0.2$ jets and large-radius $R = 1.0$ jets reclustered using small-radius anti-$k_t$ $R = 0.2$ subjets as constituents.  These large-radius jets, reconstructed as described in \Sec{sec:reconstruction}, are required to have $p_T>158$ GeV. Our analysis of jet suppression in this Section follows the analysis presented by ATLAS in Ref.~\cite{ATLAS:2023hso}.


\begin{figure} [t]
    \begin{center}
    \subfloat[$L_{\rm res} = 0$]
    {\includegraphics[width=0.55\textwidth]{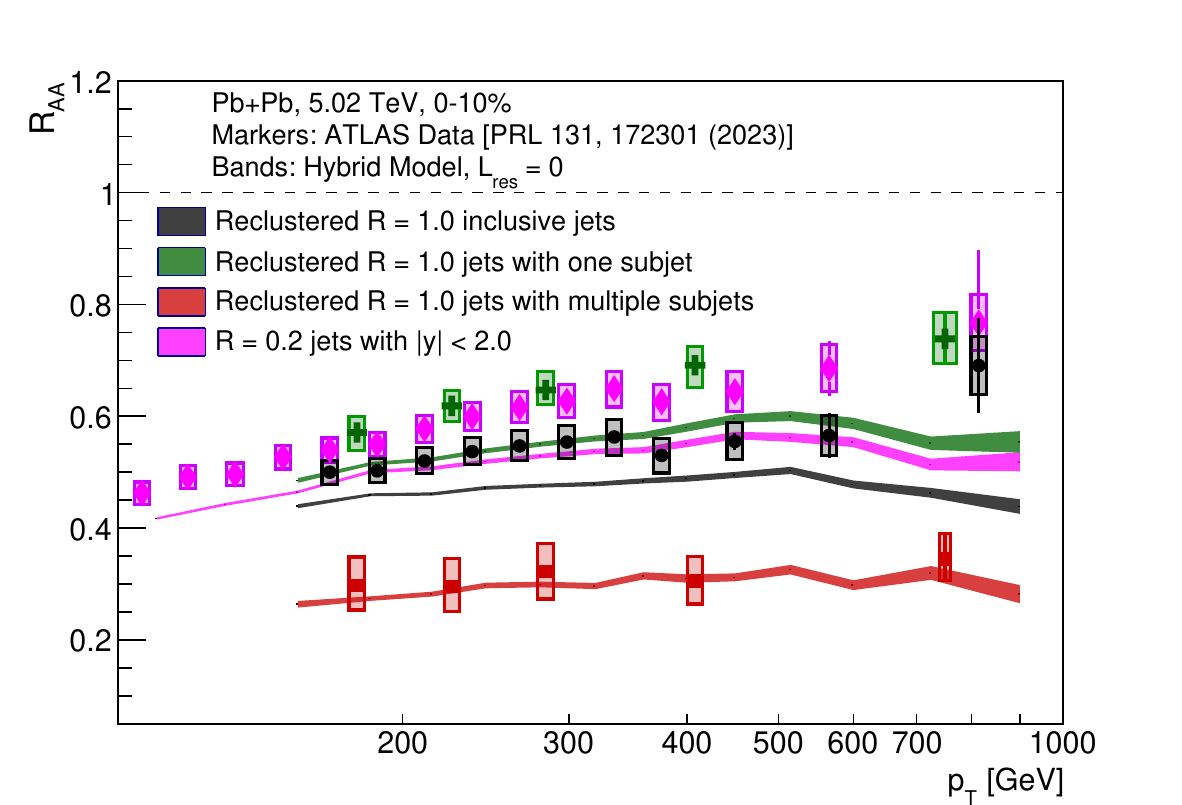}}
    \subfloat[$L_{\rm res} = 2/(\pi T)$]{\includegraphics[width = 0.55\textwidth]{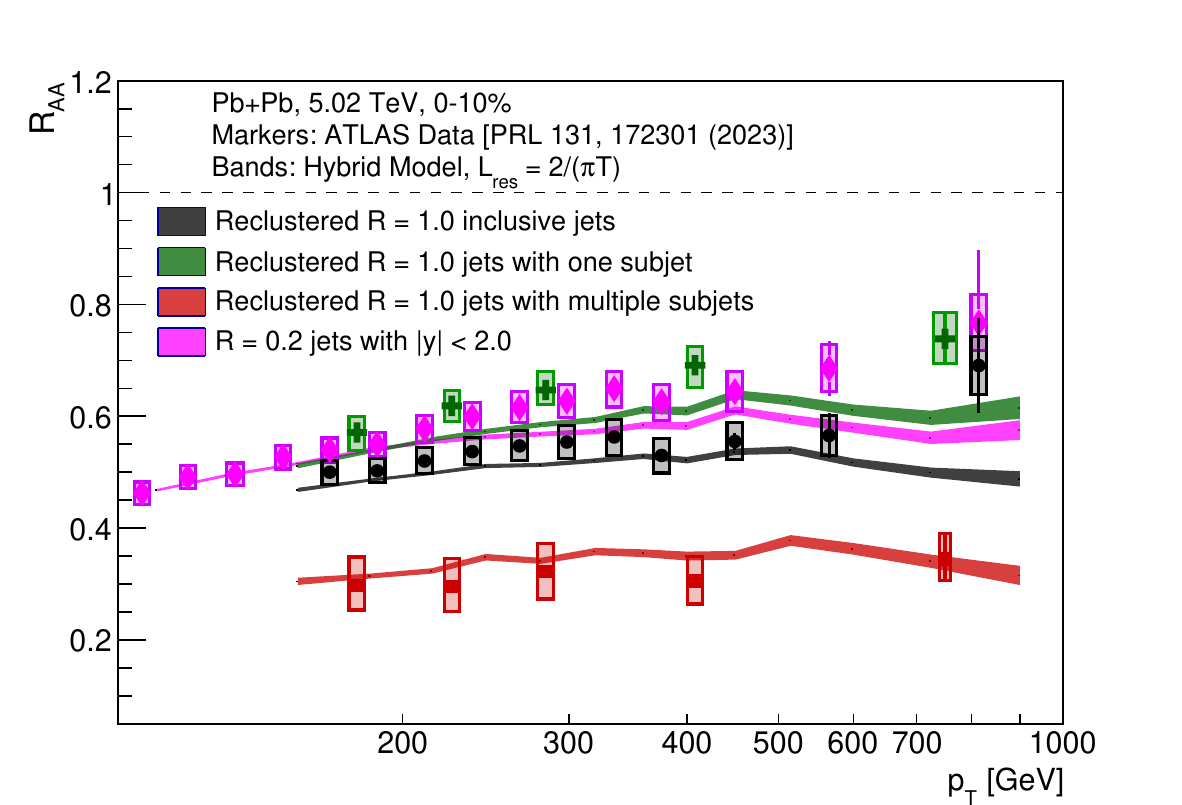}}\\
    \subfloat[$L_{\rm res} = \infty$]{\includegraphics[width = 0.55\textwidth]{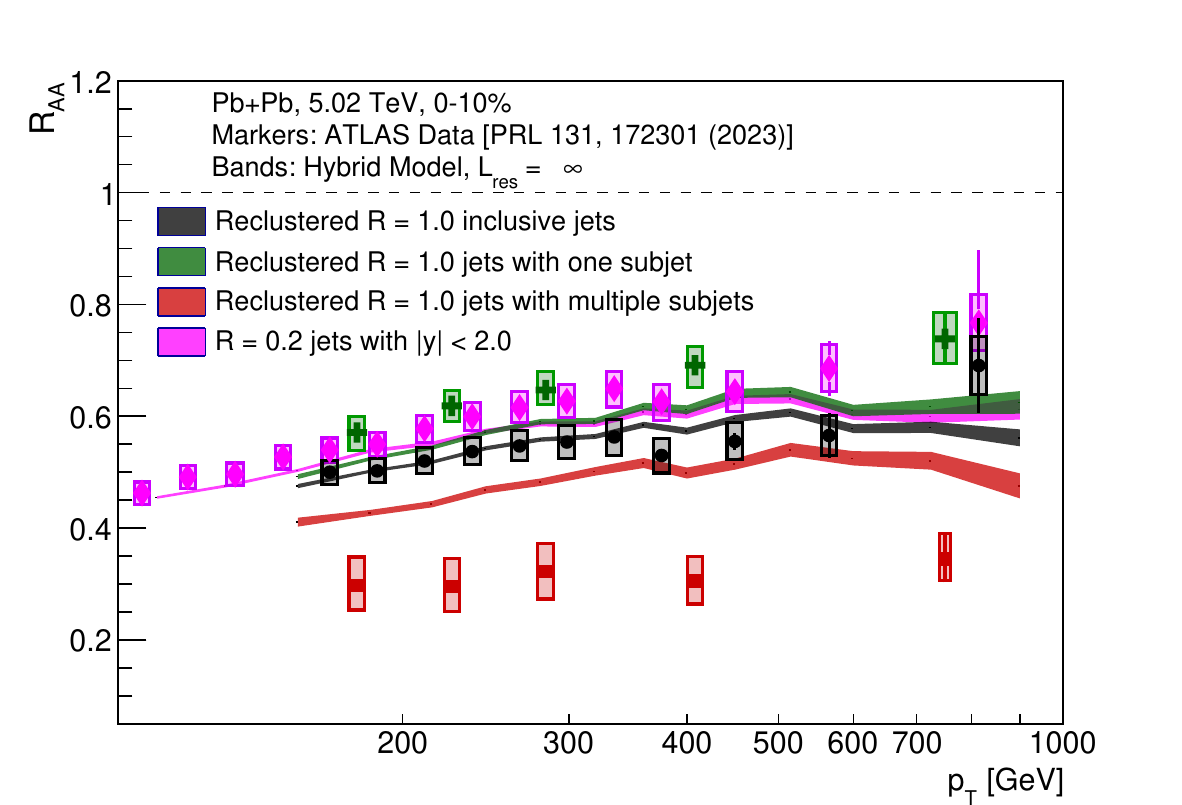}}
    \caption{Jet $R_{\rm AA}$ as a function of jet $p_T$, calculated for large-radius $R = 1.0$ jets with a single subjet (green/crosses),  $R = 1.0$ jets with multiple subjets (red/squares), $R = 1.0$ inclusive of any number of subjets (black/circles), and anti-$k_t$ $R = 0.2$ jets (pink/diamonds). Our Hybrid Model calculations, performed with (a) $L_{\rm res}=0$, (b) $L_{\rm res}=2/(\pi T)$, and (c) $L_{\rm res}=\infty$, are depicted using the colored bands. ATLAS experimental measurements from Ref.~\cite{ATLAS:2023hso} are depicted using point markers. The vertical bars on ATLAS' experimental data points indicate statistical uncertainties, while the shaded boxes indicate systematic uncertainties~\cite{ATLAS:2023hso}.}
    \label{fig:pt}
    \end{center}
\end{figure}

We begin in \Fig{fig:pt} by studying jet $R_{\rm AA}$, the suppression of jet production with a given $p_T$ in PbPb collisions relative to that in pp collisions, as a function of jet $p_T$ for anti-$k_t$ $R = 0.2$ skinny jets (pink) and for large-radius $R = 1.0$ jets with one subjet (green), multiple subjets (red), and inclusive in any number of subjets (black). Each subfigure shows Hybrid Model calculations of $R_{\rm AA}$ upon assuming different values of the QGP resolution length. We compare our results to ATLAS data from Ref.~\cite{ATLAS:2023hso}. 

The first thing to observe in the ATLAS data is the large measured difference between the suppression of large-radius $R = 1.0$ jets with multiple subjets (red) and those of large-radius $R = 1.0$ jets with a single subjet (green) or the anti-$k_t$ $R=0.2$ jets (pink). This behavior is reproduced reasonably well by the Hybrid Model with $L_{\rm res}=0$ or $L_{\rm res}=2/(\pi T)$, as shown in panels (a) and (b) of \Fig{fig:pt}. In our model, this behavior simply stems from the fact that (as long as the QGP resolution length is finite)
large-$R$ jets with multiple subjets (red) have a larger number of independent sources of energy loss than those with a single subjet (green) or than inclusive skinny jets samples (pink). In contrast, in panel (c) we see that choosing  $L_{\rm res}=\infty$ fails to reproduce experimental data, yielding too little suppression for the multiple subjets sample (red).


It is at first surprising to observe that in the case where $L_{\rm res} = \infty$, we still see that large-radius jets containing multiple subjets are suppressed slightly more than large-radius jets containing a single subjet, even though when $L_{\rm res} = \infty$ all the partons within a single  shower lose energy coherently to the plasma, as a single object. The explanation for this small effect is that two subjets that end up clustered together within a single $R = 1.0$ jet may originate from two separate parton showers. When this happens, in the Hybrid Model the two subjets are assumed to lose energy independently 
regardless of the value of $L_{\rm res}$. In Appendix~\ref{app:isr}, we show that large-radius $R=1.0$ jets may contain subjets originating from initial state radiation in addition to those originating from the hard scattering. This explains the slightly  
greater suppression of large-radius jets containing multiple subjets even when $L_{\rm res}=\infty$.


Our Hybrid Model calculations 
with $L_{\rm res}=0$ and $2/(\pi T)$ 
furthermore show that, as in the ATLAS measurements, large-radius $R = 1.0$ jets composed of only a single anti-$k_t$ $R = 0.2$ subjet (green) are the least suppressed; in fact, they are suppressed slightly less than a standard sample of anti-$k_t$ $R = 0.2$ jets (pink) which can include some events in which there are additional jets near the selected $R=0.2$ jet.
We see from the comparison in \Fig{fig:pt} that the pattern of ordering of the suppression (red most suppressed, then black, then pink, green least suppressed) is the same as that seen in ATLAS data. We shall return to the quantitative differences between Hybrid Model calculations and ATLAS data further below.

\begin{figure}[t]
    \begin{center}
    \includegraphics[width=0.8\textwidth]{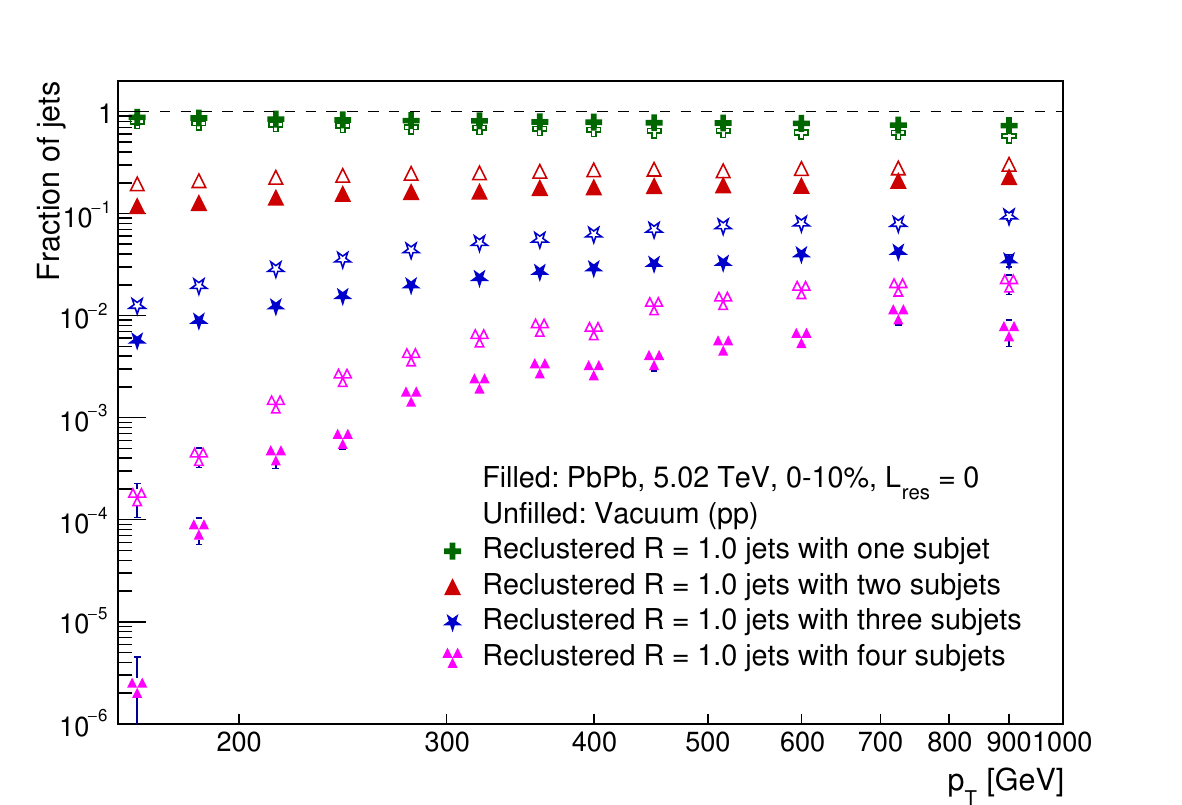}
    \caption{Fraction of large-radius $R = 1.0$ jets with one subjet (green pluses), two subjets (red triangles), three subjets (blue stars), and four subjets (pink triangle-triplets), calculated in vacuum (unfilled points) and in PbPb collisions using the Hybrid Model with zero QGP resolution length (filled points).}
    \label{fig:multiplicity}
    \end{center}
\end{figure}

In \Fig{fig:multiplicity}, we have plotted the fraction of large-radius $R = 1.0$ jets with different numbers of skinny subjets in our Hybrid Model calculations with  $L_{\rm res} = 0$ for PbPb collisions, as well as in pp collisions, as a function of the transverse momentum of the large-radius $R = 1.0$ jet. We note that the fraction of large-radius $R = 1.0$ jets with multiple subjets increases as a function of large-radius jet-$p_T$ in both PbPb and vacuum (pp collisions). This means that, on average, there are more sources of energy loss within more energetic large-radius jets in PbPb collisions, mildly reducing $R_{\rm AA}$ with increasing jet $p_T$. 
This behavior counters the mild increase of jet $R_{\rm AA}$ with $p_T$ that is characteristic of inclusive single-subjet structures, thus
yielding an overall rather flat dependence of $R_{\rm AA}$ with jet $p_T$ for these large-radius jets. 
This explanation was suggested by ATLAS in~\cite{ATLAS:2023hso}, which we have now corroborated using our model calculations with $L_{\rm res}=0$.
Fig.~\ref{fig:multiplicity} also explains why the results for large-radius large-$R$ jets with any number of subjets (black) lies closer to the results for large-$R$ jets with one subjet (green) than those with multiple subjets (red): most of the large-$R$ jets contain only one subjet.  

Although the pattern of the ordering of the suppression in the four observables shown in Fig.~\ref{fig:pt} is the same in the Hybrid Model with either $L_{\rm res}=0$ or $L_{\rm res}=2/(\pi T)$ as in the ATLAS measurements~\cite{ATLAS:2023hso},
there are certainly quantitative differences between our Hybrid Model results for the various $R_{\rm AA}$'s and the experimental measurements. These differences 
are reduced upon choosing $L_{\rm res}=2/(\pi T)$ rather than $L_{\rm res}=0$, especially in the lowest jet $p_T$ range. 

Although the comparisons that we have made certainly demonstrate that infinite $L_{\rm res}$, which is to say a picture in which entire parton showers lose energy coherently as single objects, is incompatible with the ATLAS data, 
and although we see a modest preference for $L_{\rm res}=2/(\pi T)$ over $L_{\rm res}=0$,
we shall not attempt to determine the optimal value of $L_{\rm res}$ in the Hybrid Model. 
Doing so would require a quantitative investigation of the tradeoffs in different (ideally many different) observables 
coming from increasing $L_{\rm res}$
(which shifts $R_{\rm AA}$ upwards for jets with multiple subjets) 
and from increasing $\kappa_{\rm sc}$ (which shifts $R_{\rm AA}$ downwards for all jets but moreso for jets with multiple subjets).
The value of $\kappa_{\rm sc}$ 
that we are employing was obtained from what was a global fit to extant data at the time it was 
done~\cite{Casalderrey-Solana:2018wrw}, 
but this global fit was dominated by single hadron suppression data 
which had the smallest experimental uncertainties at the time this fit was performed.  And, single hadron suppression has very little sensitivity to
the value of $L_{\rm res}$. 
%
%
Another input that impacts jet $R_{\rm AA}$, especially at very high jet $p_T\sim 1$ TeV, is the initial state effects incorporated in our model via the nPDFs that we take from the central values of the fits performed in 
Ref.~\cite{Eskola:2009uj}. It has more recently been shown that initial state effects can result in values of $R_{\rm AA}\sim 0.8$ even in the absence of quenching~\cite{Pablos:2019ngg,Huss:2020dwe,Caucal:2020uic,Adhya:2021kws,Pablos:2022mrx}. 
%
%
Thus, quantifying the constraints on $L_{\rm res}$ implied by comparisons between Hybrid Model calculations and experimental data  requires refitting the value of $\kappa_{\rm sc}$ as well as incorporating the uncertainties in the nPDFs.
Doing so would require looking at many more experimentally measured observables than we shall look at in this paper, certainly including jet and single hadron suppression but also including many well-measured jet substructure observables, in all cases for many centrality classes and, ideally, for both RHIC and LHC kinematics and collision systems.
This should be done as a Bayesian study along the lines that have been pursued in other contexts in Refs.~\cite{Bernhard:2016tnd,Nijs:2020roc,JETSCAPE:2021ehl,Heffernan:2023utr,JETSCAPE:2024cqe,Falcao:2024zkw}, and in the best case the constraints coming from the full suite of data would suffice to determine the optimal choices for 
Gaussian broadening~\cite{Casalderrey-Solana:2016jvj} and Moli\`ere scatterings~\cite{Hulcher:2022kmn}\footnote{We caution that simultaneously incorporating a nonzero QGP resolution length, or color coherence effects, with the physics of elastic scatterings including the dynamics of the associated recoiling particles remains a non-trivial theoretical challenge, and has to date not been attempted in any model of jet quenching. For first steps in this direction, see Ref.~\cite{Pablos:2024muu}.}
in the Hybrid Model, as well as for $\kappa_{\rm sc}$ and $L_{\rm res}$.
We leave this to future work.

The very substantial difference in $R_{\rm AA}$ between large-radius $R = 1.0$ jets containing only a single skinny subjet and large-radius $R = 1.0$ jets composed of multiple skinny subjets motivates
exploring the dependence of the large-radius jet suppression on the angular distribution of the skinny subjets 
within the
$R = 1.0$ jets. 
In order to define an observable that characterizes the angular separation among the skinny subjets, we begin by reclustering 
the constituents of the large-radius $R = 1.0$ jets using the $k_t$ recombination algorithm, which tends to combine the hardest constituents last. By looking at the final step of the $k_t$ reclustering, we gain access to the separation in angle $\Delta R_{12} \equiv \sqrt{\Delta y_{12}^2 + \Delta \phi_{12}^2}$ among pairs of skinny subjets that corresponds to the hardest splitting in the large-radius jet.
Studying $R_{\rm AA}$ for such reclustered large-radius $R = 1.0$ jets as a function of $\Delta R_{12}$ enables us to study the dependence of jet suppression on the angular separation between the hard structures within
large-radius jets. The skinny subjets employed in this analysis introduced by the ATLAS collaboration are energetic and skinny enough that their production must originate from hard splittings in the DGLAP evolution of the parton shower, even though they have subsequently evolved within the QGP medium.

\begin{figure}
    \begin{center}
    \includegraphics[width=0.8\textwidth]{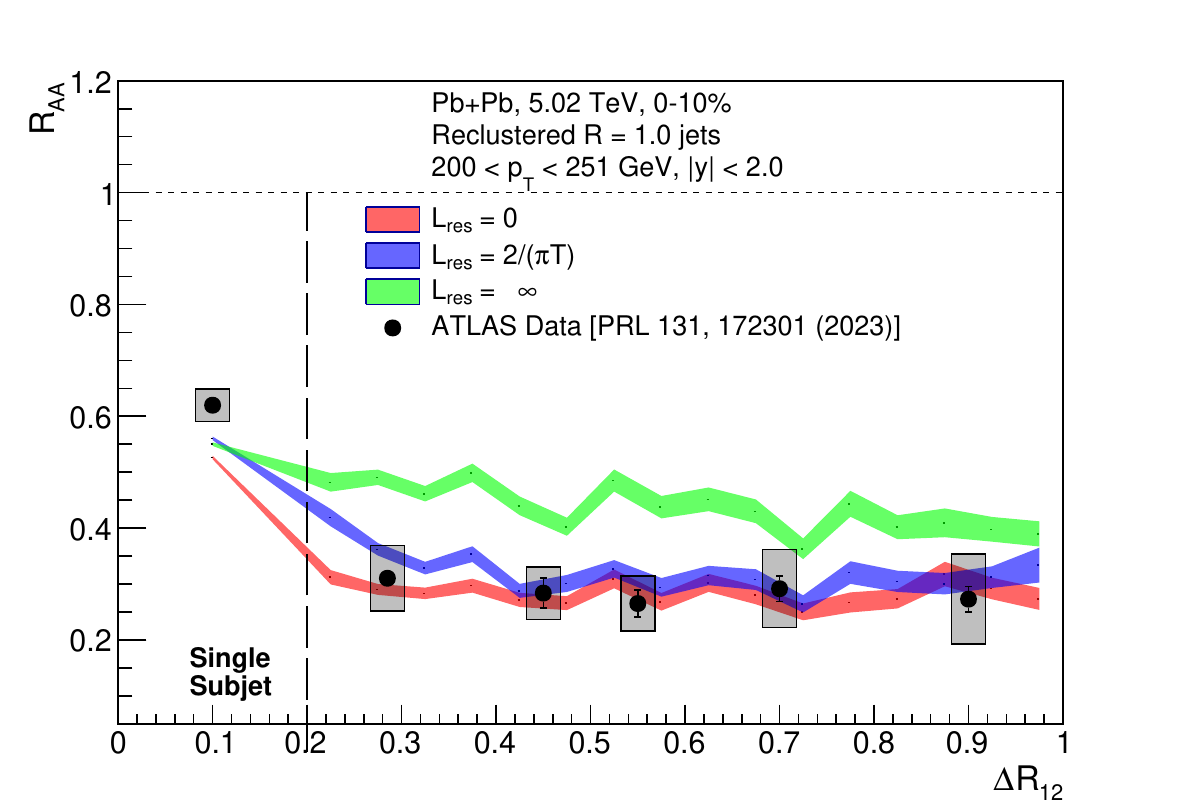}
    \caption{$R_{\rm AA}$ as a function of $\Delta R_{12}$ for reclustered large-radius $R = 1.0$ jets with $200 < p_T < 251$ GeV. The left-most bin denotes the value of $R_{\rm AA}$ for $R = 1.0$ jets that contain only a single skinny subjet, placed arbitrarily at 0.1; all bins with $\Delta R_{12}\geq 0.2$ show $R_{\rm AA}$ for $R=1.0$ jets composed of multiple skinny subjets with the hardest splitting given by $\Delta R_{12}$, as described in the text. The three colored bands show the results of Hybrid Model calculations with $L_{\rm res}=0$ (red), $2/(\pi T)$ (blue) and $\infty$ (green).
    ATLAS experimental measurements from Ref.~\cite{ATLAS:2023hso} are depicted using point markers. The vertical bars on ATLAS' experimental data points indicate statistical uncertainties and the shaded boxes indicate systematic 
    uncertainties~\cite{ATLAS:2023hso}.}
    \label{fig:deltaR}
    \end{center}
\end{figure}

In \Fig{fig:deltaR} we present the results from our Hybrid Model calculations of jet suppression that are differential in $\Delta R_{12}$ for  large-radius $R = 1.0$ jets upon assuming three different values for the QGP resolution length: $L_{\rm res} = 0$, $2/(\pi T)$, and $\infty$. 
We compare the results of our calculations against ATLAS data~\cite{ATLAS:2023hso}, obtaining good agreement when $L_{\rm res} = 0$ or $2/(\pi T)$. Consistent with our results for $R_{\rm AA}$ as a function of jet $p_T$ shown in Fig.~\ref{fig:pt}, we see that when $L_{\rm res} = 0$ the yields of  $R = 1.0$ jets with multiple skinny subjets are suppressed by a factor of $\sim 2$ more than those of large-radius $R = 1.0$ jets containing only a single skinny 
subjet, regardless of the separation $\Delta R_{12}$ between the hardest splitting in the $R = 1.0$ jet. 
When $L_{\rm res} = 0$ and when there are multiple skinny subjets within the large-radius jet, $R_{\rm AA}$ is flat as a function of $\Delta R_{12}$ because the medium is able to resolve every skinny subjet regardless of how far apart they are in angle. 
In other words, since for $L_{\rm res} = 0$ each parton in the jet shower loses energy to the plasma independently at every time step in the shower evolution, the energy loss experienced by partons in the jet shower is independent of the separations between these partons. 
Note that this {\it does not} mean that what happens to the lost energy and momentum -- how it is redistributed in angle and $p_T$ via the evolution 
of the wake(s) that the partons create in the droplet of QGP -- is independent of the angular separation between the hard structures within the jet.
Addressing this question will be the main focus of \Sec{sec:substructure}. 

The QGP resolution length characterizes the way that jets with different hard fragmentation patterns are perceived and quenched by the medium.
When the QGP has a nonzero resolution length $L_{\rm res}=2/(\pi T)$, we see that $R_{\rm AA}$ is not flat as a function of $\Delta R_{12}$
over the range $0.2<\Delta R_{12} \lesssim 0.4$.
This is because energetic partons in the jet shower that are separated by a distance less than $L_{\rm res}$ will lose energy to the plasma as if they were one unresolved parton and it becomes more likely that the QGP cannot resolve the hard partons from which the skinny subjets originate the smaller the value of $\Delta R_{12}$. 
Beyond $\Delta R_{12}\gtrsim 0.35$, the Hybrid Model results for $L_{\rm res}=0$ and $L_{\rm res}=2/(\pi T)$ agree with each other. 
This reflects the fact that above a certain angular separation, given roughly by $\theta_{\rm res} \simeq 2/(\pi T L)$, with $L$ the typical length traversed by a hard parton in the shower between one splitting and the next, the hard substructures will be resolved, and quenched separately, in both scenarios. 
This suggests that $\theta_{\rm res}\approx 0.35$ is not unreasonable, as for example it could correspond 
to 
$L=2$ fm and $T=0.2$ GeV.
While today's experimental measurements with the current uncertainties shown in Fig.~\ref{fig:deltaR} are consistent 
with our Hybrid Model calculations with
either $L_{\rm res}=0$ or $L_{\rm res}=2/(\pi T)$, it is at the same time clear from Fig.~\ref{fig:deltaR} that future measurements of this observable have the potential to give us direct access to constraints on the QGP resolution length. 
It would be very interesting to extend
the angular resolution of this measurement 
down to skinny subjets separated only by $\Delta R_{12} \sim 0.1-0.2$ where the distinction between the values of $R_{AA}$ obtained upon assuming
$L_{\rm res}=0$ or $L_{\rm res}=2/(\pi T)$ is greatest.
This would require repeating the analysis using skinnier subjets with $R=0.1$. Being able to access this lower angular range using skinny subjets will complement the information provided by jet substructure measurements performed by ALICE~\cite{ALargeIonColliderExperiment:2021mqf,ALICE:2023dwg}, ATLAS~\cite{ATLAS:2022vii} and CMS~\cite{CMS:2024zjn,CMS:2024koh,CMS:2024msa} using grooming techniques~\cite{Larkoski:2014wba} and different recombination schemes~\cite{Bertolini:2013iqa}.

We see from Fig.~\ref{fig:deltaR} that choosing
$L_{\rm res} = \infty$
is inconsistent with the experimental data, consistently with what is seen in Fig.~\ref{fig:pt}.
Choosing
$L_{\rm res} = \infty$
means that entire parton showers 
lose energy as if the originating parton had never split; the medium is completely unable to resolve the hard structures within a parton shower.
This at first suggests that the green band in Fig.~\ref{fig:deltaR}
should be completely flat, with the same suppression for large-radius $R=1.0$ jets that contain any number of skinny subjets with any angular distribution.  We see from the Figure that the green band is indeed much flatter than the blue and red bands, making it inconsistent with the data. However, the green band is not completely flat, with the $R_{\rm AA}$ for large-radius $R = 1.0$ jets with multiple skinny subjets being somewhat lower than the $R_{AA}$ for those containing only subjet, and declining slightly with increasing 
$\Delta R_{12}$. 
In the Hybrid Model with 
$L_{\rm res} = \infty$,
this is only possible if two subjets in the same large-radius $R = 1.0$ jet originate from different shower-initiating partons, as for example if one or both originates from initial state radiation before the hard scattering. 
In \App{app:isr}, we show that a 
significant percentage of the $R = 1.0$ jets that contain multiple subjets do indeed include subjets coming from initial state radiation. The probability of this happening increases with $\Delta R_{12}$, as was previously noted in Ref.~\cite{Zapp:2022dhq}, because initial state radiation tends to be emitted at larger rapidities. 
The slight decrease in $R_{\rm AA}$ as a function of $\Delta R_{12}$ for large-radius $R = 1.0$ jets in QGP with $L_{\rm res}=\infty$ 
can thus arise as a consequence of 
initial state radiation. 
Regardless, 
the scenario with
$L_{\rm res} = \infty$
is strongly disfavored by data.
This conclusion highlights the importance of 
understanding how the parton shower evolves in position-space and time 
if one seeks to understand how it
interacts with a spacetime-dependent medium like the expanding cooling droplet of QGP produced in a heavy-ion collision.
This conclusion also highlights that 
information about how hard partonic substructures within a parton shower each engage with the droplet of QGP can be imprinted in experimental observables -- which is a necessity if we hope to use jets as probes of the microscopic structure of QGP.

All of the results that we have reported in this Section are consistent with the assumption that those jets possessing a larger number of energy loss sources, generated by hard and semi-hard vacuum-like fragmentation, will experience more total energy loss whereas narrower jets containing fewer energy loss sources lose less energy.
Because the production cross section for jets is a steeply falling function of jet-$p_T$, there is a strong selection bias (sometimes called a survivor bias) favoring those jets with a given $p_T$ that lose the least energy. This survivor bias favors
narrower jets~\cite{Milhano:2015mng,Casalderrey-Solana:2016jvj,Rajagopal:2016uip,Brewer:2017fqy,Brewer:2018mpk,Casalderrey-Solana:2019ubu,Caucal:2020xad,Du:2020pmp,Brewer:2021hmh,Caucal:2021cfb}, effectively filtering out the wider, more active, jets in each $p_T$ bin. This by-now-standard picture is only realized if the 
medium can
resolve different sources of energy loss within a parton shower, which is to say if it can resolve hard partonic substructure within a jet. As such, this standard picture relies upon 
the QGP resolution length being finite.
It is therefore pleasing to see an explicit example in which chooosing
$L_{\rm res} = \infty$
results in predictions for an observable that are strongly disfavored by experimental data.

We have focused in this Section
on observables sensitive to differences in the energy loss, and consequent suppression, of large-radius jets with differing substructure -- comparing those consisting of only a single skinny subjet to those containing more than one, with varying angular separation.
However, these observables do not
tell us anything about what happens
to the ``lost'' energy.  It is of course not actually lost: it is transferred from the jet to the medium, becoming a wake in the droplet of QGP.
In the next Section we ask how the differences between the wakes of large radius jets with varying substructure
is imprinted upon the soft cloud of hadrons that is known to accompany jet quenching phenomena.

\section{Jet and Wake Substructure} \label{sec:substructure}
\subsection{Visualizing Jets and Their Wakes}

As jets traverse the droplet of QGP produced in a heavy ion collision, they lose energy and momentum to the plasma. These energy and momentum depositions excite hydrodynamic modes -- they create a wake in the plasma that carries the momentum lost by the jet.
This means that the wake has net momentum in the jet direction.
The wake is composed of two types of hydrodynamic excitations: propagating sound modes that describe compression and rarefaction of the QGP and diffusive modes. Most
of the momentum is carried by  diffusive modes corresponding to a region of QGP behind the jet that is flowing in the jet direction, following behind the jet.
Since jet wakes are the consequence of jet energy loss in a strongly coupled liquid, they encode valuable information about the thermalization process experienced by jets as they propagate through this medium and
about the hydrodynamization and subsequent evolution of the region of strongly coupled QGP disturbed by the passage of the jet. 
Hence, their study represents a key goal in jet quenching physics.

In a droplet of QGP with zero resolution length, each parton that is formed within the medium produces its own wake according to Eq.~(\ref{eq:onebody}), as described in \Sec{sec:hybridintro}. In this Section, we examine the structure of wakes produced by large-radius jets with radius $R = 2.0$ that are reconstructed by clustering anti-$k_t$ $R = 0.2$ skinny subjets using the anti-$k_t$-recombination algorithm. In other words, they are reconstructed from skinny subjets in the same way as the reclustered $R = 1.0$ jets in the previous Section, but with an even larger radius $R = 2.0$. We examine jets with such a large radius because it gives us access to a much larger phase space to study how the internal structure of large-radius jets shape their wakes. Throughout this Section, we study jets and their wakes in $\gamma$-tagged events. The criteria for selecting events, photons, and jets that we shall employ in the analyses we describe in this Section were all given in \Sec{sec:setup}. Summarizing, the large-radius $R = 2.0$ jets are reconstructed using anti-$k_t$ $R = 0.2$ skinny subjets with $p_T^{\rm jet} > 35$ GeV and $|\eta| < 3.0$ that are $\Delta \phi_{{\rm jet}, \gamma} > 2\pi/3$ away from an isolated photon with $p_T^\gamma > 150$ GeV and $|\eta^\gamma| < 1.44$. Furthermore, we require that each $R = 2.0$ jet has $p_T > 50$ GeV and $|y| < 2.0$. Throughout this Section, we shall set $L_{\rm res}=0$ in the Hybrid Model for simplicity, as we have checked that for the observables that we shall investigate here making this choice or choosing $L_{\rm res}=2/(\pi T)$ makes little visible difference to the quantities that we shall plot and makes no difference to any conclusions.

The observable of interest to us in this Section will be the \textit{differential jet shape} (or simply \textit{jet shape}), which is a measure of the angular distribution of transverse energy within and around a jet. 
Conventionally, the jet shape is defined by the
expression
\begin{equation} \label{eq:normalshape}
    \rho(y, \phi) = \frac{1}{N_{\rm jets}} \frac{1}{\delta y \text{ } \delta \phi} \sum_{\rm jets} \frac{\sum_{i \in (y \pm \delta y /2, \phi \pm \delta \phi / 2)} p_T^{i}}{p_T^{\rm jet}},
\end{equation}
where $\{i\}$ is the set of hadrons around the jet axis (here, the axis of the large-radius jet)
and where $y$ and $\phi$ are the rapidity and azimuthal angle of hadrons as measured around the jet-axis, respectively. We have chosen to normalize the jet shape \eqref{eq:normalshape} by the number of selected jets $N_{\rm jets}$. 
Note that the hadrons entering the determination of the jet shape need not only be those hadrons which were defined as belong to the jet via the jet reconstruction algorithm.
Once the jet axis has been determined, and in an analysis of experimental data once a background subtraction has been performed, the jet shape is determined by all the hadrons around the jet axis.

The conventional definition
of the jet shape \eqref{eq:normalshape} really only makes sense if we are seeking to describe jets that are, on average, azimuthally symmetric about the jet axis. In the present context, that means that this definition is well-suited to visualize the shapes of large-radius jets composed of only a single skinny-subjet, as in this case the axis of the large-radius jet will coincide with the skinny-subjet. Our principal purpose in this Section is to extend the definition \eqref{eq:normalshape}
so as to be able to visualize the shapes of large-radius jets composed of two well-separated skinny subjets -- in which the conventional jet axis for the large-radius jet will lie somewhere between the two skinny subjets, a case for which \eqref{eq:normalshape} 
is not well-suited. We shall turn to this below in Section~\ref{sec:NewObservable}, but first we begin with some general observations that apply both to the conventional definition \eqref{eq:normalshape} and to our extension thereof and with a look at large-radius jets with only a single skinny subjet, where \eqref{eq:normalshape} is relevant.

Observe that we can define different variants of the jet shape by choosing which hadrons are included in its definition \eqref{eq:normalshape}.
For example, we can look at the jet shape composed only of those hadrons with $p_T$ below some relatively low cutoff, so as to analyze the distribution of soft particles around a jet.  Or, we can use only those hadrons in some specified intermediate $p_T$ range.
These are examples of choices that can be made either in the Hybrid Model or in an analysis of experimental data.
In the Hybrid Model, though, we have further possibilities, since here (but not in experimental data) we can identify which hadrons 
are the result of jet fragmentation (i.e.~originate from the parton shower that emerges from the droplet of QGP after losing some of its energy) and which hadrons 
originate from the hadronization of the jet wake
at freeze-out. 
It is particularly interesting to look at the jet shape determined via including only those hadrons that originate from the jet wake in the definition \eqref{eq:normalshape},
as we can do in the Hybrid Model but not in experiment, as this allows us to directly visualize the shapes of jet wakes.
We shall refer to jet shape observables determined using only hadrons originating from the wake of the jet as ``wake shapes''.
We can then investigate which choice of $p_T$-range yields an experimentally realizable jet shape observable that is a reasonable proxy for the wake shape.


\begin{figure}[t]
    \begin{center}
    \subfloat[Jet Shape]{\includegraphics[width=0.55\textwidth]{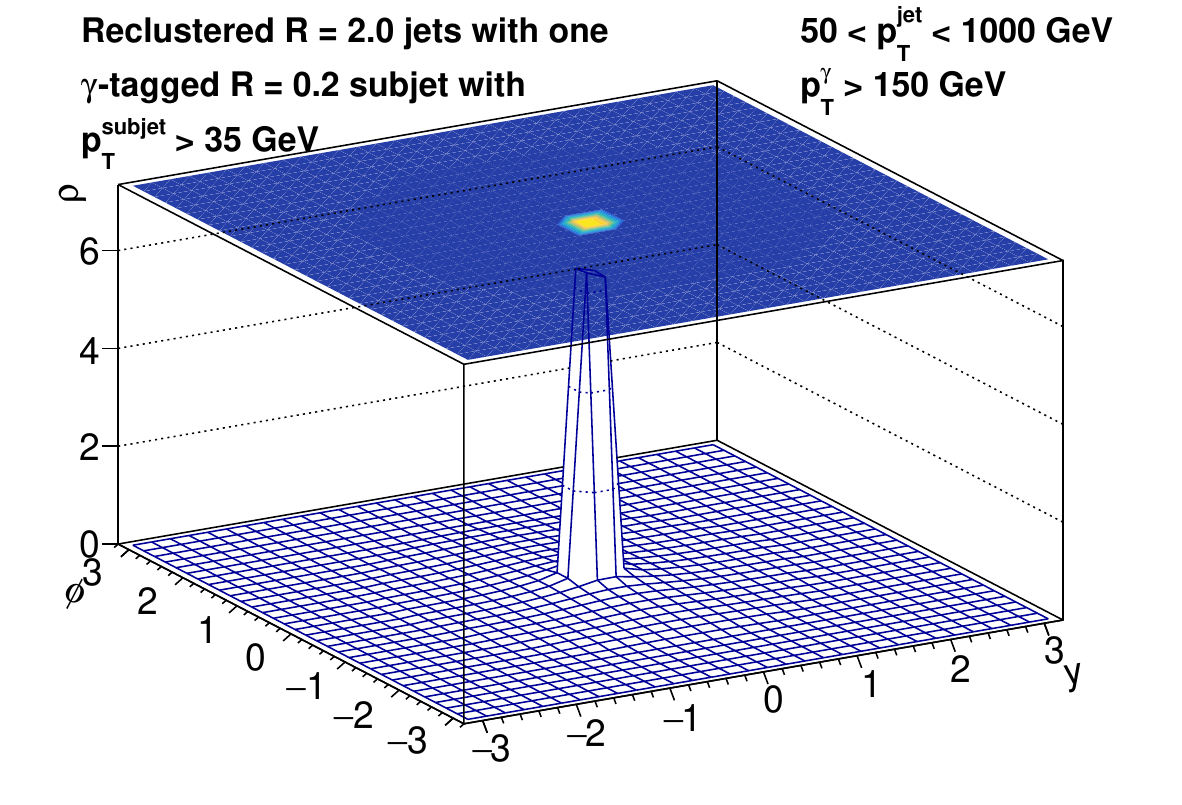}}
    \subfloat[Wake Shape]{\includegraphics[width=0.54\textwidth]{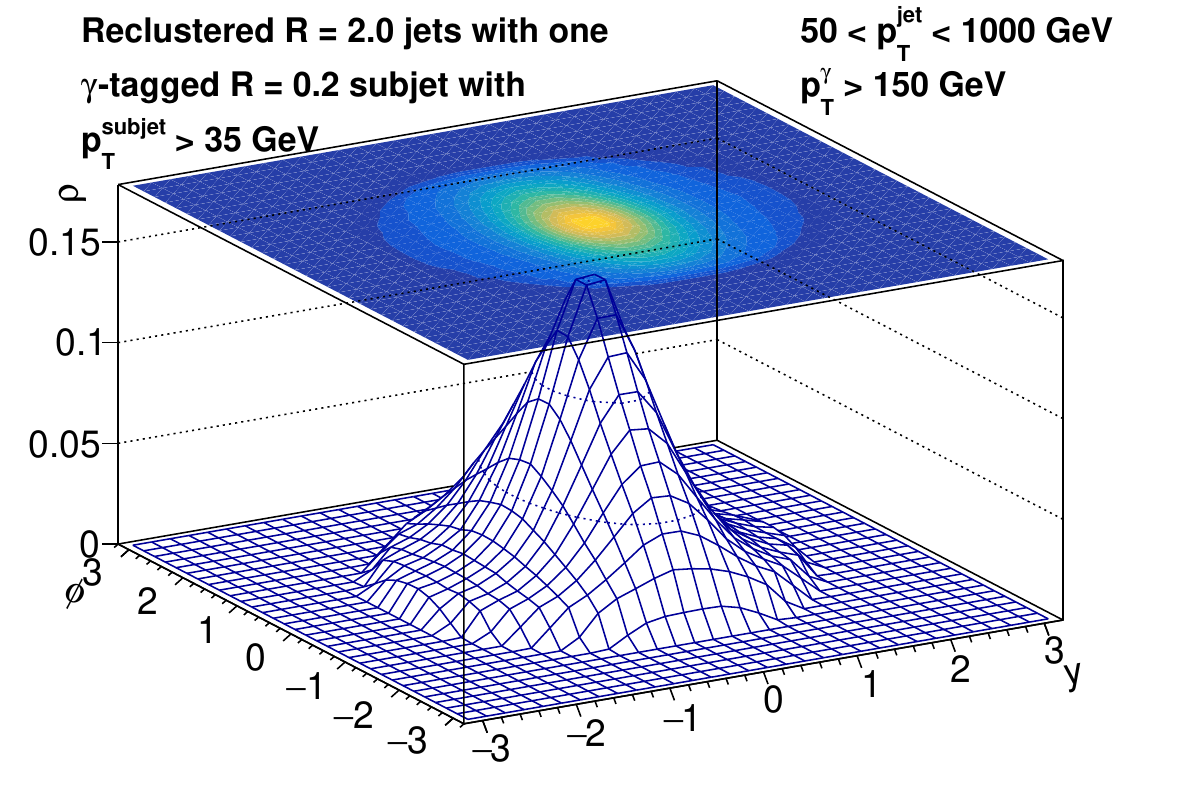}}
    \caption{Hybrid Model results for the shape of reclustered $R = 2.0$ jets (a) and their wakes (b) with one anti-$k_t$ $R=0.2$ skinny subjet in $\gamma$-tagged events, calculated using all hadrons within a radius of 2.0 around the jet axis, not just the hadrons from the skinny subjet. $y$ and $\phi$ are the rapidity and azimuthal angle respectively, as measured relative to the axis of the single skinny subjet.}
    \label{fig:ssj}
    \end{center}
\end{figure}

In \Fig{fig:ssj} we show Hybrid Model calculations of the jet shape 
for reclustered $R = 2.0$ jets consisting of only a single skinny subjet.  In panel (a) we include all
all hadrons that lie within a radius of $\Delta R = 2.0$ from the large-radius jet-axis in the determination of the jet shape. In panel (b) we show the shape of the wake produced by such jets, 
which is to say the jet shape calculated using only
those hadrons originating from jet wakes found within a radius of $\Delta R = 2.0$ around the axis of a reclustered $R = 2.0$ jet that contains only one single skinny 
subjet. 
Note the different vertical scales in the two panels; the broad wake is in fact visible in panel (a) if one looks carefully after one learns from panel (b) what to look for.
Of course, the full jet shape in panel (a) is dominated
by the narrow high peak at the origin corresponding to the single skinny subjet.
The wake produced by the hard collinear partons in the parton shower that yielded the skinny subjet is highlighted in panel (b); it is much broader and softer. 


It is important to note that had there been two roughly back-to-back jets in each event in the ensemble, as it is the case for an ensemble of inclusive jets or for an ensemble of dijet events,
then the wake shape observed in the right panel of \Fig{fig:ssj} would have been distorted by the presence of the jet(s) sitting in the opposite hemisphere. We are employing $\gamma$-jet events in the analysis of this Section to avoid this complication.
The reason why jets in the opposite hemisphere would distort the shape of the wake is that as we have discussed following Eq.~\eqref{eq:onebody}, the wake 
of each individual jet includes an excess of soft particles around the jet direction (which we are seeking to visualize here) as well as a depletion of soft particles in the opposite direction in the transverse plane relative to the uncorrelated background that is subtracted by experimentalists.
This is apparent via the 
$\cos (\phi-\phi_j)$ term 
in Eq.~\eqref{eq:onebody}, and has been seen in experimental data by CMS in $Z$-jet events~\cite{CMS:2024fli}. 
Consequently, if 
there were other jets in the opposite hemisphere from the large-radius jet, the resulting depletion of soft particles in {\it their} backward hemisphere would distort the shape of the wake of the large-radius jet that we are seeking to visualize, and that we have shown in the right panel of \Fig{fig:ssj}.
Removing this complication from our analysis is the reason why we have chosen to study an ensemble of photon-jet events in the present Section. 
Both for completeness and because we recognize that although wake shapes in an ensemble of inclusive jet events will be harder to interpret they will always have higher statistics in experimental data than ensembles of gamma-jet events, we present and discuss the corresponding results for inclusive jet ensembles like those in Sec.~\ref{sec:suppression} and those analyzed by ATLAS in Ref.~\cite{ATLAS:2023hso} in Appendix~\ref{app:inc}. It goes without saying that, learning about the interplay between the wakes generated by two back-to-back jets is also of great interest, but we defer such a dedicated study to future work.

Rather, in the present work, our focus is the interplay between two wakes sitting in the same hemisphere. We know that large-radius jets with at least two subjets are more suppressed than single-subjet ones, as the amount of energy loss is increased, on average, by roughly a factor 2, as discussed in Sec.~\ref{sec:suppression}. It is then natural to ask whether one can visualize the result of these two independent processes of energy loss by means of studying the correlated yields of soft particles around these structures. Next, we introduce new jet shape observables that are specifically designed to this end, allowing 
us to study how the structure of a large-radius jet -- specifically how the angular separation between two skinny subjets in a large-radius jet -- shapes their wakes.


\subsection{A New Jet Shape Observable}
\label{sec:NewObservable}

\begin{figure}[t] 
\begin{center}\includegraphics[width=0.4\textwidth]{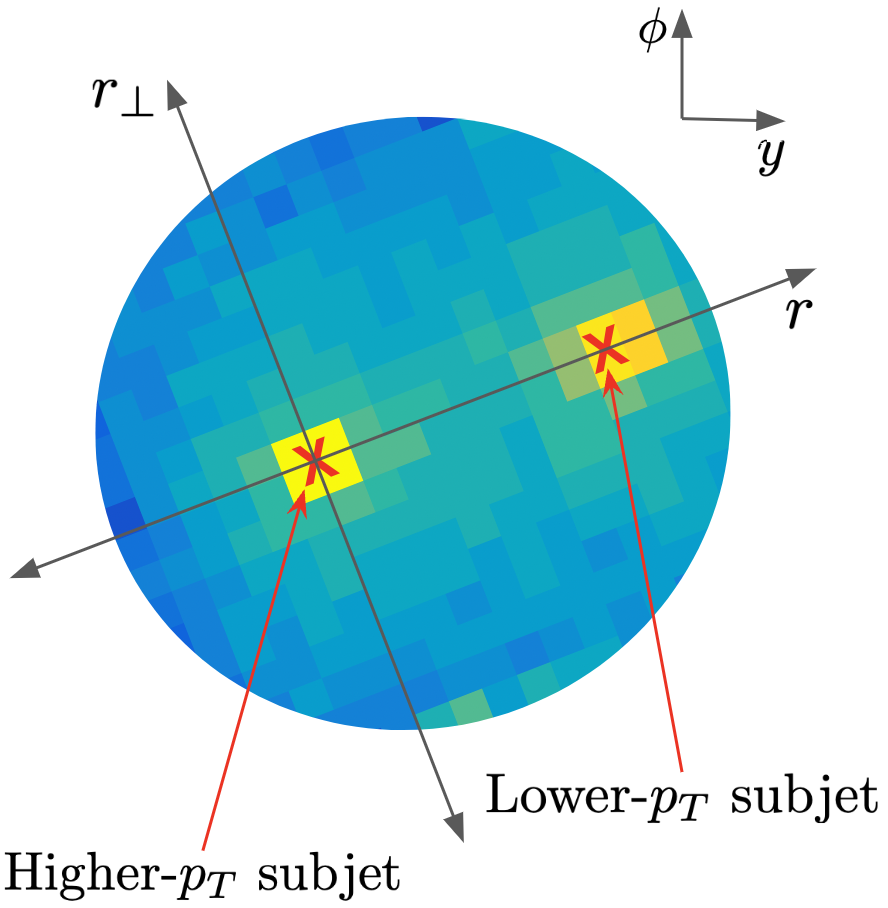}
    \caption{Heuristic depiction of the $p_T$-weighted distribution of hadrons within an $R = 2.0$ jet with two skinny subjets. The origin of the $(r,r_\perp)$ coordinates is centered on the higher-$p_T$ subjet, with the $r$-axis pointing through the center of the lower-$p_T$ subjet (see the red X's). The $r_\perp$ axis is perpendicular to the $r$-axis. Note that the origin of the $(r,r_\perp)$ coordinates is {\it not} located at the conventionally defined axis of the entire large-radius jet, as that lies somewhere between the two skinny subjets.}
    \label{substructure-shape-setup}
    \end{center}
\end{figure}

Suppose we have a large-radius jet containing $n>1$ skinny subjets. For a given one such large-radius jet, its energy distribution across the detector will feature $n$ sharp peaks, each of which is centered at the axis of the corresponding subjet. Since these subjets can lie anywhere within the large-radius jet, aggregating the jet shape of all such structures according to \Eq{eq:normalshape} would 
lose all of their most interesting features by averaging
over the $R = 2.0$ circle around the large-radius jet-axis. 
Here, we shall define a 
new jet shape observable that is suitable for $n=2$, namely for visualizing the shapes of large-radius $R=2.0$ jets reclustered from precisely two skinny anti-$k_t$ $R=0.2$ subjets.
%
We focus on $n=2$ and not larger both because it is the simplest non-trivial choice that allows us to visualize how jet substructure shapes jet wakes, but also because it enables us to construct an intuitive observable that will allow us to cleanly, directly and simply
visualize the shape of large-radius jets containing two skinny subjets. We construct such an observable as follows.
Given a large-radius $R = 2.0$ jet with two skinny subjets, we define a 2-dimensional coordinate system $(r, r_\perp)$ via the following procedure. First, let the higher-$p_T$ and lower-$p_T$ subjets be located at respective coordinates $(y^\text{high}, \phi^\text{high})$ and $(y^\text{low}, \phi^\text{low})$ in the rapidity-azimuthal angle plane. We define the origin of our new coordinate system to be at $(y^\text{high}, \phi^\text{high})$ and we define the $r$-axis to point positively in the direction of $(y^\text{low}, \phi^\text{low})$. Next, we define the $r_\perp$-axis to be perpendicular to the $r$-axis. And finally, we choose the handedness of the new coordinate system such that $\hat{r} \times \hat{r}_\perp = \hat{y} \times \hat{\phi}$. \Fig{substructure-shape-setup} shows a pictorial setup of these axes, using a heuristic histogram of a $p_T$-weighted distribution of hadrons within an $R = 2.0$ jet with two skinny subjets. The locations of the two $R = 0.2$ subjets in the Figure are identified by the red X's in the $y$-$\phi$ plane.

Note that in the new $(r, r_\perp)$ coordinates, by construction the axes of the two skinny subjets will always lie on the $r$-axis. We shall be particularly interested in examining the region between two subjets --- these coordinates are well-suited for this purpose. With these new coordinates chosen, we can now redefine the jet shape (whose conventional definition isEq.~\ref{eq:normalshape})
in the way that the illustration
in \Fig{substructure-shape-setup}
prompts us to do.  We define the value of our new more differential jet shape observable suitable for visualizing the shapes of large-radius jets containing two skinny subjets
at some location $(r, r_\perp)$ in our new coordinate system as the fraction of a jet's hadronic energy contained within a $\delta r \times \delta r_\perp$ box centered at the location $(r, r_\perp)$:
\begin{equation} \label{newjetshape}
    \rho(r, r_\perp) \equiv \frac{1}{N_{\rm jets}} \frac{1}{\delta r \text{ } \delta r_\perp} \sum_{\rm jets} \frac{\sum_{i \in (r \pm \delta r /2, r_\perp \pm \delta r_\perp / 2)} p_T^i}{p_T^{\rm jet}},
\end{equation}
where again, $i$ runs over \emph{all} the hadrons that lie within an angle $\Delta R=\sqrt{\Delta \phi^2+\Delta y^2}<2$ from the axis of the reclustered $R = 2.0$ jet, not only those hadrons within the skinny subjets. 
In the above expression, $p_T^\text{jet}$ is the $p_T$ of the reclustered $R = 2.0$ jet and, as  in Eq.~\eqref{eq:normalshape}, the expression is normalized by the number of selected reclustered $R = 2.0$ jets $N_{\rm jets}$ that pass the selection criteria, which now will include a condition on the
separation in rapidity between the skinny subjets, $\Delta y_{12}$, as we explain in the next Subsection.



\subsection{How Jet Substructure Shapes Jet Wakes}

In \Sec{sec:suppression}, we examined how large-radius jet suppression depends on the substructure of such jets via
looking at the dependence of $R_{AA}$ on the separation $\Delta R_{12} = \sqrt{\Delta y_{12}^2 + \Delta \phi_{12}^2}$ between the final two constituents involved in the hardest $k_t$-splitting of a reclustered $R = 1.0$ jet. For a reclustered $R = 1.0$ jet with precisely two skinny subjets, $\Delta R_{12}$ is just the angle between the two subjets. We now have the tools we need in order to visualize how the medium responds to the internal structure of such jets. More precisely, we can visualize how the shape of the soft hadrons originating from the freezeout of the wakes of large-radius jets containing two well-separated skinny subjets depends on the separation between the subjets. 
Although in \Sec{sec:suppression} we used $\Delta R_{12}$ to characterize the separation between subjets, in this Section we shall study the shapes of reclustered $R = 2.0$ jets with two subjets -- and of their wakes -- as a function of the separation $\Delta y_{12}$ in rapidity between the two subjets. We choose $\Delta y_{12}$ as our differential variable instead of $\Delta R_{12}$ because the wake -- as currently implemented in the Hybrid Model -- is wider in azimuthal angle $\phi$ than  in rapidity $y$. This can be seen in \Fig{fig:ssj}, and it corresponds to the shape of the $\exp{-\frac{m_T}{T} \cosh(y-y_j)}$ term in Eq.~\eqref{eq:onebody}. Hence, binning in the rapidity separation between subjets should make the appearance of the sub-wakes produced by each subjet clearer.

\begin{figure} [H]
    \begin{center}
    \subfloat[Jet Shape]{\includegraphics[width=1.05\textwidth]{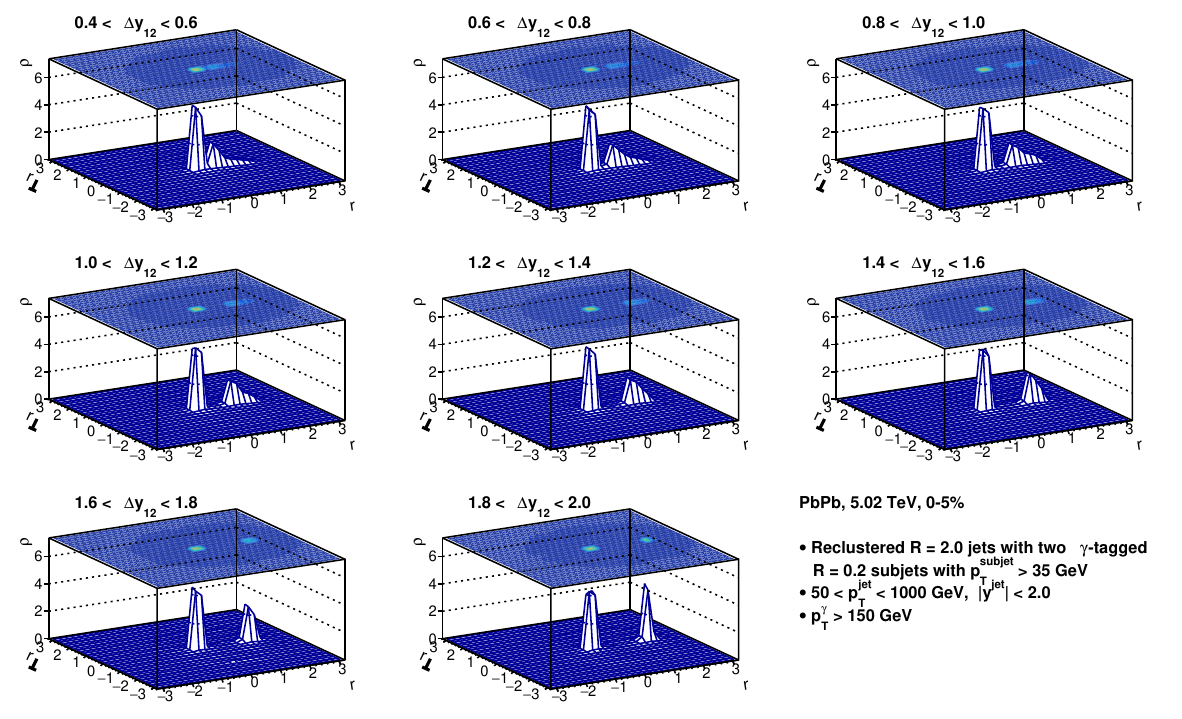}}\\
    \subfloat[Wake Shape]{\includegraphics[width=1.05\textwidth]{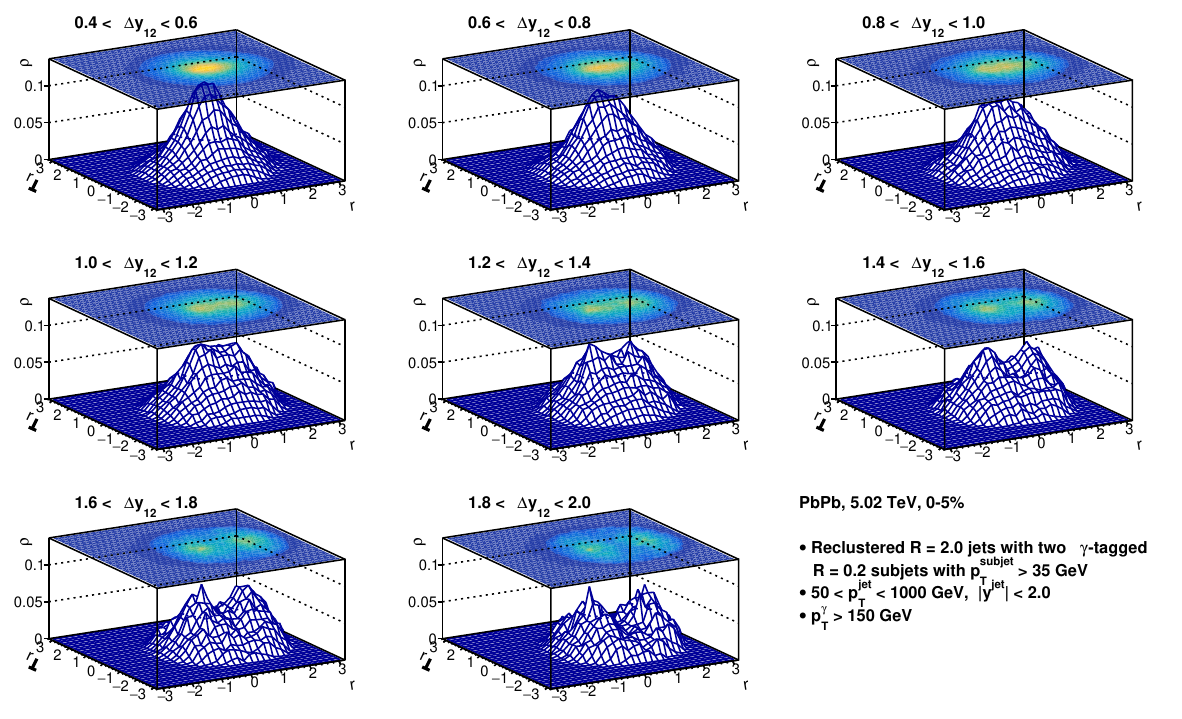}}
    \caption{Shapes of reclustered $R = 2.0$ jets (a) and their wakes (b) with two anti-$k_t$ $R=0.2$ subjets in $\gamma$-tagged events, calculated using hadrons within a radius of 2.0 around the $R = 2.0$ jet-axis. Different panels show different ranges of the separation $\Delta y_{12}$ in rapidity between the two skinny subjets.}
    \label{fig:allparticleshapes}
    \end{center}
\end{figure}

\Fig{fig:allparticleshapes} shows the shapes of the reclustered $R = 2.0$ jets with two skinny subjets (upper panels) and the shapes of their wakes (lower panels), calculated using hadrons within a 2.0 radius of the axis of the reclustered $R = 2.0$ jet. The different panels in the upper (and lower) halves of the figure each depict the jet (and wake) shapes for large-radius jets whose two skinny subjets are separated by different ranges $\Delta y_{12}$. If we first examine the jet shapes in the upper panels, we 
see what we expect visualized clearly: we see distinct peaks at the locations of 
each of the two subjets. Because the subjets are skinny anti-$k_t$ $R=0.2$ subjets, these peaks are distinct, tall, and sharp. These peaks are dominated by the hadrons coming from the fragmentation of the collinear parton showers.
In every one of the upper panels in \Fig{fig:allparticleshapes}, the two peaks corresponding to the two skinny subjets are separate and distinct.
%
We also note that in each of the upper panels the peak at the location of the lower-$p_T$ subjet, which by construction is on the right, at positive $r\sim\Delta y_{12}$, possesses a trailing edge that extends to values of $r > \Delta y_{12}$. This is because two subjets that are separated by $\Delta y_{12}$ in rapidity can also be separated by a nonzero $\Delta \phi_{12}$ in azimuthal angle; such subjets will be separated by $\Delta R_{12} = \sqrt{\Delta y_{12}^2 + \Delta \phi_{12}^2} > \Delta y_{12}$, resulting in the trailing edges beyond $r = \Delta y_{12}$ seen in the upper panels of \Fig{fig:allparticleshapes}.

Next we turn to the wake shapes visualized in the lower panels of Fig.~\ref{fig:allparticleshapes}.
Of course the vertical scale in the lower panels is much different than that in the upper panels, which are dominated by the hadrons from the parton showers in the two peaks. 
Much more interesting, though, are the strikingly different shapes seen in the lower panels. We see that as long as $\Delta y_{12}\lesssim 1.2$, the two well separated skinny subjets are enveloped by a single, common, cloud of soft hadrons originating from the wake left in the QGP by the two skinny subjets.
It is only when the separation $\Delta y_{12}$ is {\it very} large, greater than 1.2, that we begin to see separated wakes around each subjet.
We see that this new differential jet shape observable for the first time allows us to visualize the overlay of the two wakes originating from two subjets with control over their degree of overlap.
Because we have used the Hybrid Model with $L_{\rm res}=0$, we know that each parton in the parton shower loses energy to the medium and sources a wake independently.
This means that each of our well-separated skinny subjets sources
its own wake in the fluid, but the hadrons originating from each wake at freezeout have such a broad distribution in angle that when they are superposed in the final hadronic state after freezeout 
the hadrons in the wake shape sourced by two well-separated independently quenched subjets can form a single broad structure.



If the lower panels in Fig.~\ref{fig:allparticleshapes} depicted predictions for a quantity that could be measured in experiment, we could  devise ways to visualize the 
shapes of jet wakes that would highlight the  striking differences between their shapes and the shapes of the two-skinny-subjet jets that sourced them.  For example, we could define projections of the 
jet shapes that would highlight how the space between the subjets is filled in by the jet wakes.
Of course, in experimental data (unlike in the Hybrid Model) the hadrons in the final state do not come with labels that tell us which hadrons originate from the
fragmentation of a parton shower and which hadrons come from the freezeout of a jet wake.
In the next Subsection (and, with further detail, in Appendix~\ref{app:harderpt})
we shall look for -- and find --  experimentally realizable differential jet shape observables that are good proxies for the wake shapes shown in the lower panels of Fig.~\ref{fig:allparticleshapes}.
%
%
We do so by taking advantage of the fact that since jet wakes are moving QGP, the hadrons that originate from jet wakes are soft, with momenta determined via the temperature and velocity of the
droplet of QGP at freezeout. 
Restricting the jet shape observables that we have introduced such that they are composed only from hadrons with soft transverse momenta maximizes the relative contribution of hadrons originating from the wake relative to hadrons originating from parton showers.

\subsection{Towards Experimental Measurements}

In this Subsection we recompute our new jet shape observable using only soft hadrons, as can be done in analyses of experimental data.
We have selected the range $0.7<p_T<1$ GeV. This low momentum range has already been accessed by CMS in the analysis of substructure observables, such as~\cite{CMS:2021nhn}, making it clear that this is an observable that is accessible in experimental data. In fact, we understand that even a somewhat lower transverse momentum range may be accessed experimentally in the near future.

\begin{figure} [t]
    \begin{center}
    \includegraphics[width=1.05\textwidth]{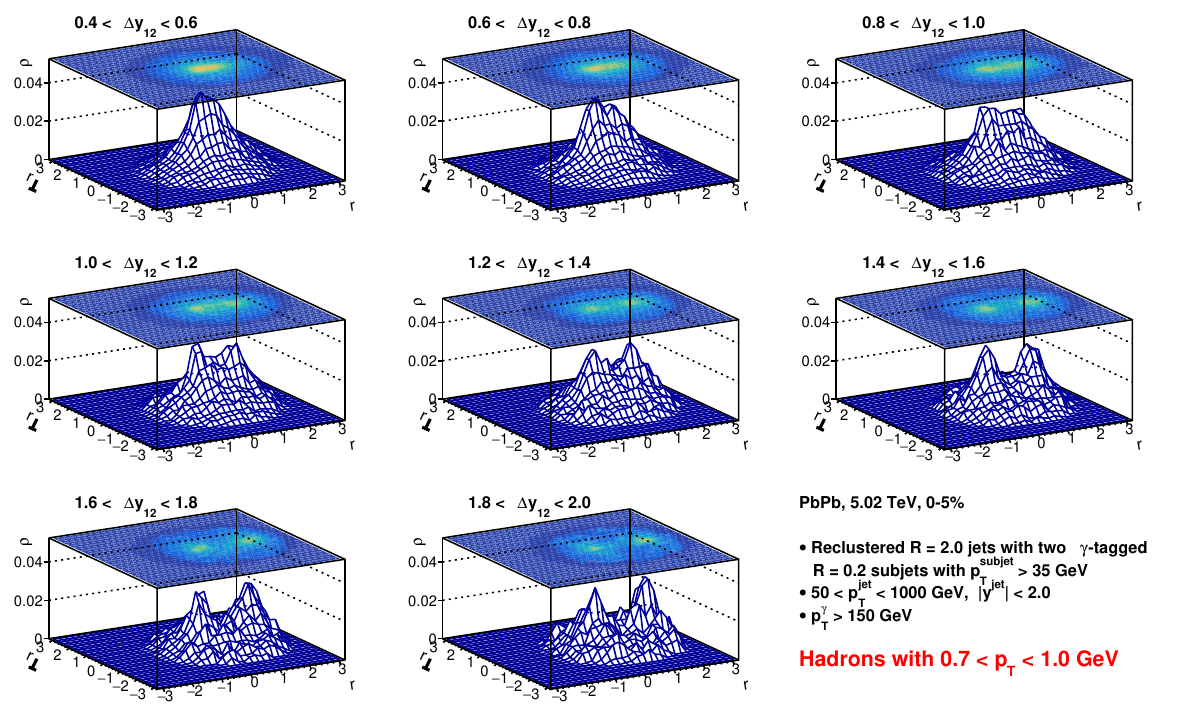}\\
    \caption{Jet shapes of reclustered large-radius $R = 2.0$ jets with precisely two anti-$k_t$ $R = 0.2$ skinny subjets in PbPb collisions, calculated using only those hadrons with $0.7 < p_T < 1.0$ GeV within a radius of $\Delta R = 2.0$ around the reclustered $R = 2.0$ jet's axis. Each panel displays our new jet shape observable constructed only from these soft hadrons for varying ranges of $\Delta y_{12}$, the separation in rapidity  between the two skinny subjets. 
        }
    \label{fig:2d-shapes-0.7-1.0}
    \end{center}
\end{figure}

In Fig.~\ref{fig:2d-shapes-0.7-1.0} we show the equivalent of the upper panels of Fig.~\ref{fig:allparticleshapes}, but restricting the $p_T$ range of
the hadrons contributing to the jet shape observable to  $0.7<p_T<1$ GeV.
These Hybrid Model calculations are comparable to those in the upper panels of Fig.~\ref{fig:allparticleshapes}
in that they depict a jet shape observable that can be measured in experimental data.
However, the jet shapes that we see
in Fig.~\ref{fig:2d-shapes-0.7-1.0}
are much more similar to 
the shapes in the lower panels
of Fig.~\ref{fig:allparticleshapes}, meaning that this jet shape
observable constructed from soft hadrons is a reasonable proxy for the wake shape.
In Fig.~\ref{fig:2d-shapes-0.7-1.0} -- as in the 
wake shapes in the lower panels
of Fig.~\ref{fig:allparticleshapes} -- we see a single, common, cloud of soft hadrons enveloping the two well-separated skinny subjets even when the two subjets are as far apart as $\Delta y_{12}\sim 1.0$.
Only for $\Delta y_{12}\gtrsim 1.2$
do we see separated soft subjet shapes.
Comparing the vertical axes (i.e.~the heights of the peaks) in 
the soft jet shapes of Fig.~\ref{fig:2d-shapes-0.7-1.0} to those of the
wake shapes in the lower panels
of Fig.~\ref{fig:allparticleshapes},
we see though that by choosing a $p_T$ range as low as we have
we are leaving out a sizeable fraction of the 
hadrons originating from jet wakes.
Even though including harder particles in the observable would mean including more of the wake, it also means including more hadrons originating from the parton shower and thus diminishes the  relative contribution of the hadrons originating from the wake to the observable.  We investigate this in Appendix~\ref{app:harderpt}, and conclude that the $p_T$-range that we have employed here yields a jet shape observable that is a reasonable proxy for the wake shape.

\begin{figure}
    \begin{center}
    \includegraphics[width=1.05\textwidth]{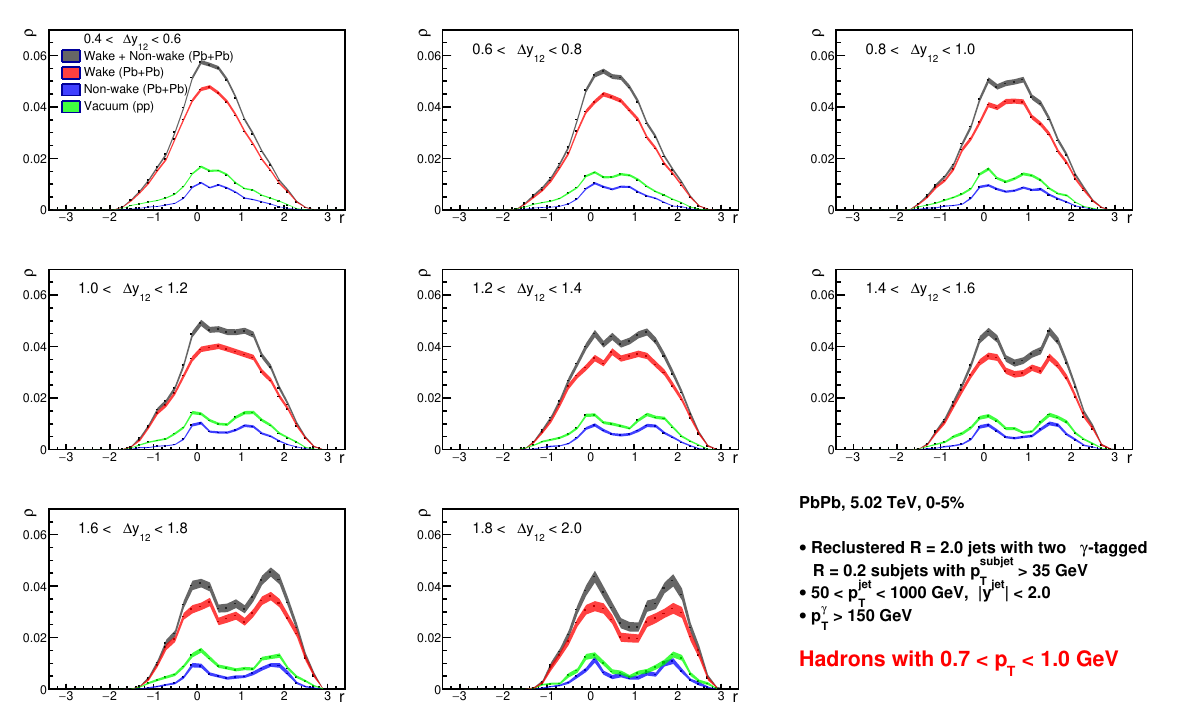}
    \caption{Projections of the 2-dimensional jets shapes of reclustered $R = 2.0$ jets with two skinny subjets onto the $r$-axis, calculated using only those hadrons with $0.7 < p_T < 1.0$ GeV within a radius of $\Delta R = 2.0$ around the reclustered $R = 2.0$ jet's axis. The bands correspond to jet shape projections calculated using only jet wake particles in PbPb collisions (red), non-wake particles from jet fragmentation in PbPb collisions (blue), all particles from jets in PbPb collisions (gray), and particles from jets in vacuum (green). Each panel corresponds to a given range of rapidity-separation $\Delta y_{12}$ between the two skinny subjets.}
    \label{fig:1d-shapes-0.7-1.0}
    \end{center}
\end{figure}

In order to facilitate the background subtraction needed in experimental analyses and to enable the extraction of quantitative conclusions via comparison between our calculations and data, it is advantageous to look at one-dimensional projections of
the full $(r,r_{\perp})$ profiles that we have plotted in Figs.~\ref{fig:allparticleshapes} and~\ref{fig:2d-shapes-0.7-1.0}.
In \Fig{fig:1d-shapes-0.7-1.0},
we perform projections onto the $r$ axis, looking at the transverse energy density profile in slices that are perpendicular to the vector linking the subjets. 
The gray bands (spread is due to statistics) correspond to large-radius jets containing two skinny subjets in PbPb collisions, while the green ones are for jets with the same selection criteria in vacuum, i.e.~in pp collisions. In the Hybrid Model (but not in experimental data) we can dissect the contributions to this observable for jets in PbPb collisions  from hadrons originating from jet wakes
relative to the contribution 
from hadrons originating from parton showers.
The contributions from only the jet wake particles are in red, and those coming from only the fragmentation of parton showers (non-wake) are in blue. (There is only a single relevant production mechanism in pp collisions, namely jet fragmentation.)
The first thing we observe is that the wake contribution (in red) represents the lion's share of the full distribution (in gray). This confirms our expectation based upon
Figs.~\ref{fig:allparticleshapes} and~\ref{fig:2d-shapes-0.7-1.0}:
by restricting the particles in the analysis to those with soft momenta $0.7<p_T<1.0$~GeV, we obtain an experimentally realizable observable that is a good proxy for just looking at particles from jet wakes.  This observable gives us a chance to 
visualize the dynamics of the wake directly, in experimental data. The second thing we note is that the non-wake PbPb (in blue) and pp contributions (in green) present a similar shape. This is expected given their common dynamical origin from the fragmentation of parton showers. 
Given that the full result is dominated by the wake contribution, it is not only until rather high values of $\Delta y_{12}\gtrsim 1.4$ that two peaks can be recognized in the gray projected distributions. This is consistent with what we see for the wake particles alone here, in the red projected distributions, and is consistent  
with what we concluded via comparing the wake shape plots in the bottom panel of Fig.~\ref{fig:allparticleshapes} to Fig.~\ref{fig:2d-shapes-0.7-1.0}. Indeed, even though the soft hadrons from jet fragmentation (non-wake) 
already present two peaks at $\Delta y_{12}\lesssim 0.8$, it is not until $\Delta y_{12}\gtrsim 1.4$ that separate structures appear for the full soft hadron distribution. 

Our investigations show that even though in all cases we have chosen $\Delta y_{12}$ large enough that
the two skinny subjets in the large-radius jets selected in our analysis are well separated, after quenching, the wakes excited by these skinny structures are so wide that they form a single common cloud of soft hadrons enveloping both subjets and only become distinguishable as two separate wakes when one imposes a rapidity separation between the skinny subjets that is very wide indeed.
Analyzing objects with angular separations this large has been recently done in experiments, such as in the ATLAS study of photon plus multi-jet correlations~\cite{ATLAS:2023wzj}. This study in fact features kinematic selections very similar to the ones we have used throughout this Section, demonstrating the statistical feasibility of the 
measurements we have proposed here even in $\gamma$-jet events. 
Regarding our choice of selecting pairs of jets based on their rapidity separation, instead of the standard combination of azimuthal angle and rapidity, current detector acceptances ($|y|<2.5$) are enough to span the largest separations that we have employed. Moreover, future detector upgrades in ATLAS~\cite{ATLAS:2021yvc}, CMS~\cite{CMS:2017jpq} and ALICE~\cite{ALICE:2022wwr} will further increase the acceptance and allow for even better statistics.

\section{Concluding Remarks and a Look Ahead} \label{sec:conclusions}

In this work we have studied several aspects of the energy loss processes experienced by large-radius jets composed of skinny subjets in heavy-ion collisions, as reconstructed using the method pioneered by ATLAS~\cite{ATLAS:2023hso}. This method consists in finding small-radius anti-$k_t$ jets first, and using them as constituents to reconstruct large-radius jets.

In Section~\ref{sec:suppression}, we have studied the degree of suppression of the large-radius jets as a function of their substructure, namely as a function of the number of skinny subjets that they contain and the separation between the skinny subjets. To do so, we have used the same jet selection criteria chosen by ATLAS in Ref.~\cite{ATLAS:2023hso}, namely employing ensembles of 
inclusive jets and reconstructing $R=1.0$ jets from skinny subjets selected with anti-$k_t$ $R=0.2$.
We have compared the results obtained with the Hybrid Model against the different jet $R_{\rm AA}$ measurements performed by ATLAS in Fig.~\ref{fig:pt}. We have found that the relative suppression as a function of jet $p_T$ among anti-$k_T$ jets with $R=0.2$, large-radius jets with $R=1$ with one subjet, two or more subjets, and any number of subjets is very well reproduced by the model, provided that the medium is able to resolve the various color charges within the large-$R$ jet. 
The Hybrid Model reproduces this pattern of suppression well if we choose the QGP resolution length $L_{\rm res}$ to be 0 or $2/(\pi T)$,
perhaps slightly better for the 
latter choice.  It completely fails to describe the data if we choose $L_{\rm res}=\infty$, which corresponds to treating entire parton showers coherently as  single sources of energy loss.
The fact that large-radius jets are more suppressed the larger the number of skinny subjets they possess, as seen in Fig.~\ref{fig:pt} both in data and in our model, directly reflects the fact that total energy loss scales with the number of resolved quenched partonic structures. 
The reason why large-$R$ jets with multiple subjets still are more suppressed than single subjet large-$R$ jets in the $L_{\rm res}=\infty$ case is due to the fact that some subjets are in fact produced via initial state radiation (see App.~\ref{app:isr}).
The modest, albeit visible improvement of the description of data achieved when going from $L_{\rm res}=0$ to $L_{\rm res}=2/(\pi T)$ motivates performing a Bayesian study to determine the optimal values for $L_{\rm res}$, $\kappa_{\rm sc}$, as well as the parameters governing  other physical effects not included in the present work (namely Gaussian broadening and Moli\`ere scattering).

Next, we have studied how the degree of suppression of large-radius jets depends on the angular separation between the two final branches in the $k_T$ clustering algorithm, which is to say on the angular separation between most separated skinny subjets. 
We have made this comparison for different values of $L_{\rm res}$ and compared the results to experimental data in Fig.~\ref{fig:deltaR}. Both $L_{\rm res}=0$ and $L_{\rm res}=2/(\pi T)$ are able to describe the fact that single subjet configurations are roughly half as suppressed as multiple subjet configurations. Moreover, the flatness of jet suppression as a function of the angular separation for multiple subjet configurations is also well reproduced, within current experimental uncertainties, for both choices of the QGP resolution length. The main differences lie at relatively small angular separations, which motivates accessing that region in the future so as to be able to better constrain the value of $L_{\rm res}$. The fully unresolved case, with $L_{\rm res}=\infty$, where only the total charge of the entire parton shower is perceived by the medium fails to describe the data. This is consistent with what we saw in  Fig.~\ref{fig:pt} and highlights the important role played by jet substructure fluctuations in the description of jet quenching phenomena.

In Section~\ref{sec:substructure}, we have visualized and analyzed the outcome, the product, the consequence, of the suppression of these large-radius jets with different spacing between their constituent skinny subjets. While $R_{\rm AA}$ quantifies the absence of jets with a given $p_T$ due to energy loss, other  observables, which are generalizations of the jet shape, enable the analysis of the fate of that ``lost'' energy and momentum, energy and momentum transferred from the jets to the plasma.
The hydrodynamic wakes excited in the droplet of QGP by a passing jet, which is to say  the jet-induced perturbations to the energy-momentum tensor of the QGP, hadronize at the freezeout hypersurface in the form of soft particles, with a boosted thermal momentum distribution and with a relatively wide angular distribution. While the excess of soft particles at large angles with respect to the jet axis has been well documented in the literature, in this work we have performed the first analysis of the overlapping shape of two of such wakes, the wakes of the skinny subjets within our large radius jets. 
The setup that ATLAS introduced and that we have employed
allows us to control the angular separation between the centers of those wakes, by selecting large-radius jets with two skinny subjets with different angular separations. 
Our analysis shows that this setup gives experimentalists the opportunity to see distinctive evidence for the presence, and shape, of jet wakes.

Instead of the ensembles of inclusive jets used in Section~\ref{sec:suppression}, in Section~\ref{sec:substructure} we have decided to use semi-inclusive photon-jet ensembles. The reason is that we wish to focus on the study of the overlap of two wakes that lie in the same hemisphere without having to deal with the 
modifications to these structures 
caused by the wakes of one or more additional jets in the opposite hemisphere, as would be present with an inclusive jet selection.
In order to be able to select large-radius jets with large angular separations between subjets, we have first selected skinny jets that recoil from an energetic photon and combined them within a large-radius $R=2$ jet. 
We have then defined a new two-dimensional jet shape observable, as is mandatory if we wish to study the shape of the soft particles from the wake around two well separated skinny subjets.
We specify the observable using two axes, 
one of which is oriented along the vector that connects the centers of the two skinny subjets. In Fig.~\ref{fig:allparticleshapes} and in Fig.~\ref{fig:1d-shapes-0.7-1.0} we have presented the results of Hybrid Model predictions for this new jet shape observable for large-radius jets with different rapidity separations $\Delta y_{12}$ between their subjets (rapidity instead of total angle, since the wakes are narrower in $y$ than in the azimuthal angle $\phi$). 
While we have selected jets consisting of two distinct, well separated, skinny subjets, 
we see that if we include all particles in the new jet shape observable, regardless of their dynamical origin, those soft hadrons coming from the wake form a strongly overlapping common cloud around the two skinny subjets even when their separation $\Delta y_{12}$ is rather large.  We see distinct, separated, wakes only when this separation is larger than 1.2-1.4.
Promisingly, we show in Fig.~\ref{fig:1d-shapes-0.7-1.0} that by restricting the momentum range of the included particles to lie within (an  experimentally accessible) range of soft momenta $0.7<p_T<1$ GeV, the new jet shape observable is dominated by the contribution from the wake and serves as a good proxy for looking only at the wake (as can be done in simulations but not in data). An experimental measurement with this setup, and within this momentum range, will give us a direct handle on the dynamics of two jet-induced wakes with different degrees of overlap.

The analytic, but oversimplified description of the hadrons coming from the wake used in this work, encapsulated in Eq.~(\ref{eq:onebody}), is known to yield hadrons that are somewhat too soft (too many hadrons with $p_T$'s well below 1 GeV; few hadrons with $p_T$'s around 2 GeV or so) and somewhat too widely distributed in angle when compared to what would obtain in a more complete treatment of jet wakes~\cite{Casalderrey-Solana:2016jvj,Casalderrey-Solana:2020rsj}. Including the 
effects of the radial flow of the droplet of QGP on the jet wakes therein (effects that are not included in Eq.~(\ref{eq:onebody}))
will, on average, collimate and harden the distribution of hadrons originating from the wake. 
This, together with the fact that our proposed observable involves two sourcing skinny subjets that can have any relative orientation relative to the outward radial flow of the droplet of QGP,
provides strong motivation for
future work on developing a computationally efficient event-by-event description of jet wakes in the flowing fluid produced in heavy ion collisions.
A study of radial flow in the new observable that we have proposed must be left to future work.

These observations confirm that the Hybrid Model calculations in the present paper cannot (currently) be taken as {\it quantitative} predictions
for, say, the value of the angular separation 
$\Delta y_{12}$ between skinny subjets beyond which two well separated wakes can be observed.
The qualitative picture, though, presents experimentalists with a striking opportunity.
An experimental measurement of our new differential jet shape observable for $0.7<p_T<1.0$ GeV hadrons around two skinny subjets separated by $\Delta y_{12}$ should first reveal a common cloud of soft particles around the two distinct subjets, and should then reveal the emergence of two separate wakes only when $\Delta y_{12}$ is increased further.
Perhaps it will only need to be increased to $\sim 1$ instead of $\sim 1.4$ to see this, time will tell.  Experimentalists thus have the opportunity to, first, see a distinctive dependence of these feature that would be a striking confirmation of the presence of jet wakes, and, second, to make a quantitative determination of the angular separation needed to see separate wakes which would give us crucial information about the degree to which radial flow boosts and collimates the hadrons from jet wakes.

In Section~\ref{sec:suppression} we have used our model studies to respond to ATLAS data, showing that their extant measurements rule out a picture in which entire parton showers are seen by the medium as single coherent sources of energy loss. Confirming that the QGP can resolve structure within a parton shower is a crucial step toward the use of jets as probes of the microscopic structure of QGP.  Furthermore, we have shown that current data mildly favors a small nonzero QGP resolution length $L_{\rm res}=2/(\pi T)$ over $L_{\rm res}=0$
and have shown how measurements to come can further constrain the value of $L_{\rm res}$.
In Section~\ref{sec:substructure}, on the other hand, we have proposed a new observable, one built upon the selection strategy introduced by ATLAS, one that is experimentally feasible, but one that has not yet been measured. This new observable presents experimentalists with the striking opportunities described above. 

We close by noting that the preliminary results recently presented by the CMS collaboration on the direct measurement of the effects of medium response using $Z$-hadron correlations~\cite{CMS:2024fli} represent long-sought unambiguous experimental confirmation of the existence of medium response effects due to the passage of the jet through the liquid QGP. 
What CMS has seen is the depletion of soft hadrons around the $Z$ originating from the 
wake(s) of jet(s) in the hemisphere opposite to the $Z$.
This crucial finding provides complementary evidence for the fluid nature of the QGP as well as for the modification of jets as they pass through this fluid and provides direct evidence for the modification of the droplet of fluid as jets pass through it.
What we have proposed in Section~\ref{sec:substructure} represents a natural step forward: the study of the phenomenology of the hydrodynamic wake(s) sourced by two separate skinny subjets.
We have provided new tools and observables 
with which to analyze the shape of multiple wakes in an experimentally realizable setup. Indeed, the core jet analysis techniques employed here have already been successfully used by experimental collaborations at the LHC. The measurement of the proposed observables in the near future will reveal the dynamics of medium response in the presence of multiple interfering sources, allowing us to better constrain the physics of jet, and their hydrodynamic wakes, in strongly coupled QGP.

\acknowledgments
We thank Carlota Andr\'es, Hannah Bossi, Brian Cole, Rithya Kunnawalkam Elayavalli, Jamie Karthein, Riccardo Longo, Jean Du Plessis,  Ian Moult, Ananya Rai, Martin Rybar, and Rachel Steinhorst  for useful discussions. Research supported by the U.S. Department of Energy, Office of Science, Office of Nuclear Physics under grant Contract Number DE-SC0011090. 
DP is funded by the European Union's Horizon 2020 research and innovation program under the Marie Sk\l odowska-Curie grant agreement No 101155036 (AntScat), by the European Research Council project ERC-2018-ADG-835105 YoctoLHC, by the Spanish Research State Agency under project 
PID2020-119632GB-I00, by Xunta de Galicia (CIGUS Network of Research Centres) and the European Union, and by Unidad de Excelencia Mar\'ia de Maetzu under project CEX2023-001318-M. ASK is supported by a National Science Foundation Graduate Research Fellowship Program under Grant No. 2141064, 
and was supported by a Euretta J. Kellett Fellowship, awarded by Columbia University. KR is grateful to the CERN Theory Department for hospitality and support.

\appendix

\section{Effects of Initial State Radiation}
\label{app:isr}

One might ask why reclustered $R = 1.0$ jets with multiple subjets are more suppressed than those jets with a single subjet in a plasma with infinite resolution length in Fig.~\ref{fig:pt}. After all, a plasma with infinite resolution length should see an entire parton shower, including all substructure within it, as a single source of energy loss, i.e. as if the parton that produced that shower had never split. We therefore hypothesize that the increased suppression of reclustered $R = 1.0$ jets with multiple subjets can be attributed to the fact that our prescription for energy loss via $L_{\rm res}$ only applies to those partons that belong to the same shower. In particular, in the Hybrid Model two partons that belong to showers with two different initiators will lose energy as independent sources, no matter how close together they are. So, a reclustered $R = 1.0$ jet with multiple subjets that each originated from a different shower-initiating partons will have more sources of energy loss than a reclustered $R = 1.0$ jet with multiple subjets that 
originated from
successive splittings within the same parton shower. Hence, the former of these two $R = 1.0$ jets will lose more energy than the latter (which will lose energy in a plasma with $L_{\rm res}=\infty$ as if the initiating parton had never split into multiple subjets).

Since hard scatterings initiate jet showers that are roughly back-to-back, it is unlikely that a reclustered $R = 1.0$ jet would contain subjets that originate from different parton showers coming from the hard scattering.
There must be another source that produces subjets which do not originate from the hard scattering. One such source is initial state radiation (ISR), which refers to the emission of quarks and gluons from the initial partonic states of the colliding nuclei prior to the moment of hard scattering.

ISR that is emitted before the hard scattering can fragment, traverse the droplet of QGP formed in the collision, and be clustered into jets during jet reconstruction. Within our treatment of the QGP resolution length described in Sec.~\ref{sec:lres}, (sub)jets that are initiated by ISR lose energy independently from (sub)jets that were initiated by the partons involved in the hard scattering (and independently from sub(jets) that were produced via other ISR), regardless of distance between the constituents of these (sub)jets' parton showers. So, in a plasma with infinite resolution length, we expect a reclustered $R = 1.0$ jet with multiple subjets where at least one of its subjets was produced by ISR to be more suppressed than a reclustered $R = 1.0$ jet with multiple subjets where none of its subjets were produced by ISR.

\begin{figure} [t]
    \begin{center}
    \subfloat[Fraction of re-clustered $R = 1.0$ jets]{\includegraphics[width = 0.54\textwidth]{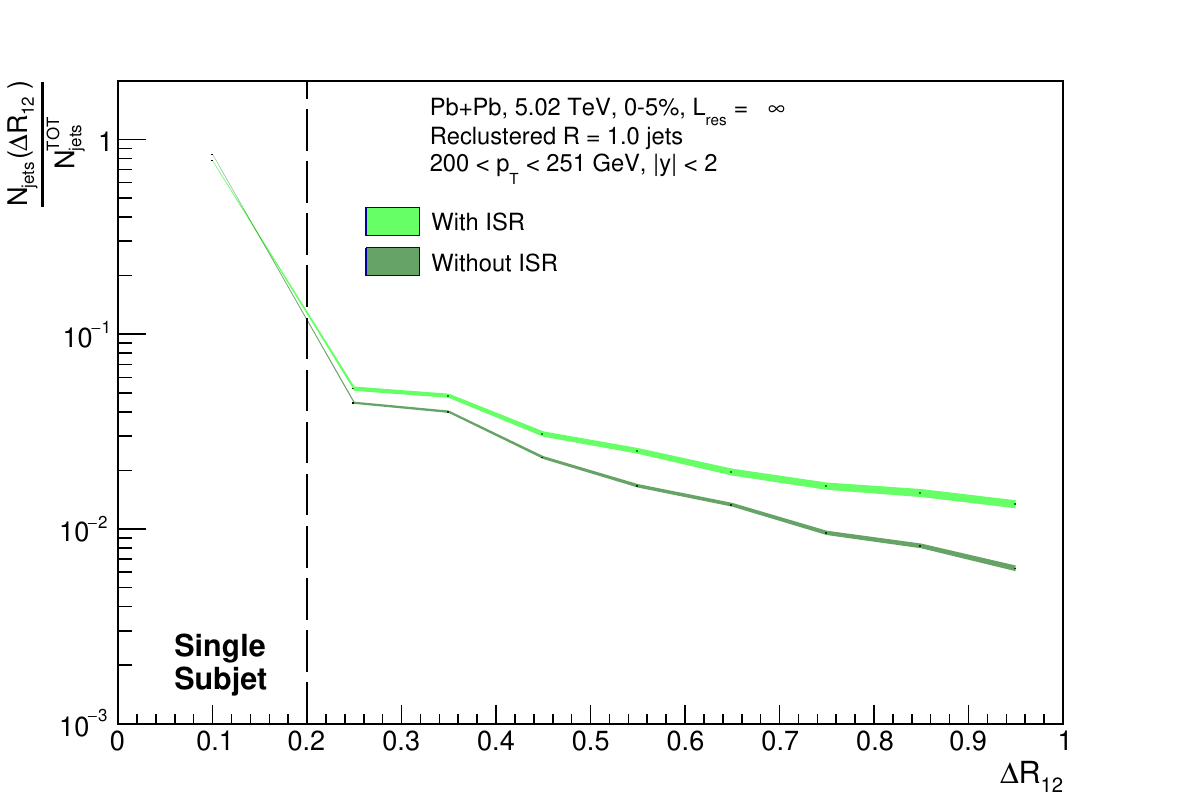}}
    \subfloat[$R_{\rm AA}$ as a function of $\Delta R_{12}$]{\includegraphics[width = 0.54\textwidth]{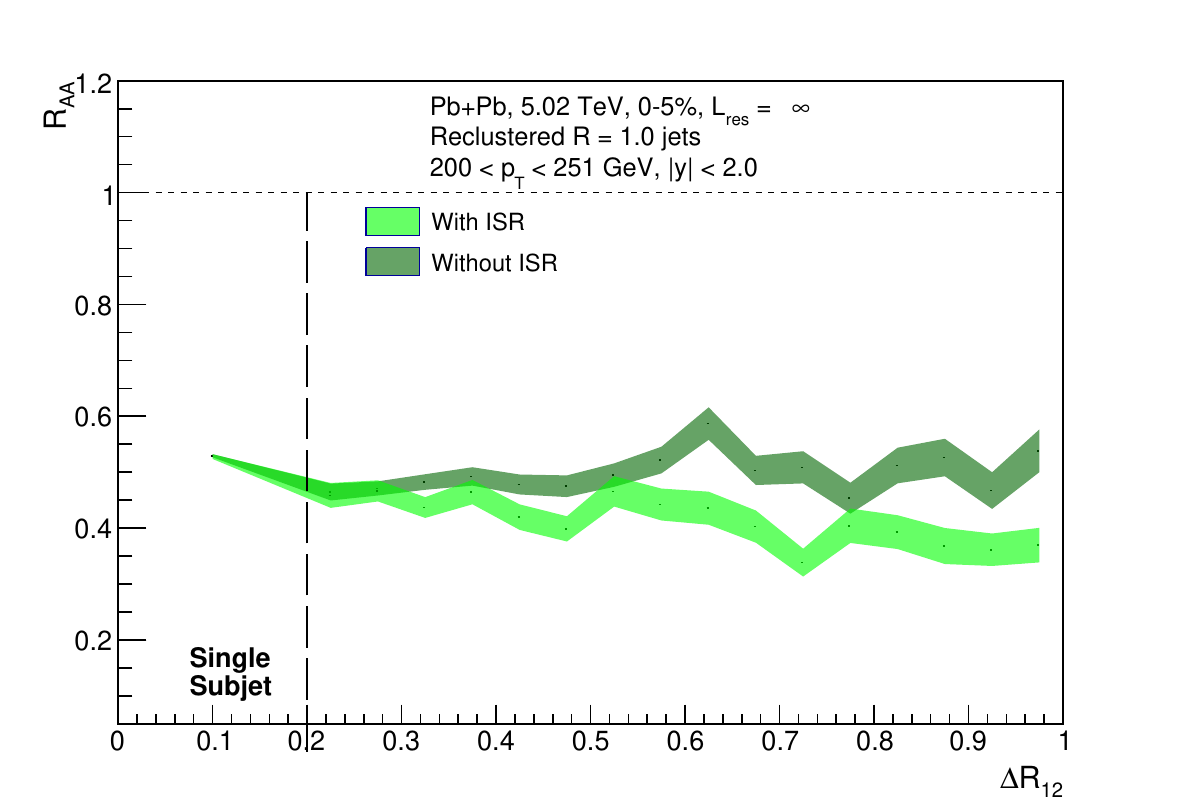}}
    \caption{(a) Fraction of re-clustered $R = 1.0$ jets with $200 < p_T < 251$ GeV contained within each $\Delta R_{12}$-bin in a sample of Pb+Pb events with $L_{\rm res} = \infty$, with ISR (light green) and without ISR (dark green). (b) $R_{\rm AA}$ as a function of $\Delta R_{12}$ for re-clustered $R = 1.0$ jets with $200 < p_T < 251$ produced in a samples of pp and Pb+Pb events (with $L_{\rm res} = \infty$) with ISR (light green) and without ISR (dark green).}
    \label{fig:isr}
    \end{center}
\end{figure}

Fig.~\ref{fig:isr}(a) shows the fraction of reclustered $R = 1.0$ jets with $200 < p_T < 251$ GeV contained within each $\Delta R_{12}$-bin produced in a sample of PbPb events where $L_{\rm res} = \infty$. The light green curve denotes the fraction of such jets produced in events with ISR. The dark green curve denotes the fraction of such jets produced in events without ISR. We note that the fraction of re-clustered $R = 1.0$ jets with multiple subjets ($\Delta R_{12} > 0$) is higher in the sample of events with ISR than in the sample of events without ISR. We also note in Fig.~\ref{fig:isr}(a) that within a given $\Delta R_{12}$-bin, the fraction of reclustered $R = 1.0$ jets containing ISR-initiated subjets increases as a function of $\Delta R_{12}$. This increase in the fraction of large-radius jets with ISR-initiated subjets at higher $\Delta R_{12}$ is supported by a similar analysis using JEWEL in Ref.~\cite{zapp_2022}, which showed that the fraction of reclustered $R = 1.0$ jets with ISR-initiated subjets increases as a function of $\sqrt{d_{12}} \equiv \rm{min}(p_{T,1}, p_{T, 2}) \frac{\Delta R_{12}}{R}$, the splitting scale of the constituents in the final $k_t$ clustering step of the $R = 1.0$ jet.

Since the probability to have a large-$R$ jet with ISR subjets increases with $\Delta R_{12}$, one expects that the $R_{\rm AA}$ of reclustered $R = 1.0$ jets decreases as a function of $\Delta R_{12}$ in a sample with ISR, even when $L_{\rm res}=\infty$. We confirm this via plotting the light green curve in Fig.~\ref{fig:isr}(b). In the absence of ISR all subjets are likely produced by successive splittings of the same initiating parton in the hard scattering -- such a collection of subjets acts as a single source of energy loss as the subjets traverse a plasma with $L_{\rm res} = \infty$. Thus, in the absence of ISR and in a plasma with $L_{\rm res} = \infty$, $R_{\rm AA}$ is constant as a function of $\Delta R_{12}$ up to statistical fluctuations, as demonstrated by the dark green curve in Fig.~\ref{fig:isr}(b). 

The results presented in the two panels of Fig.~\ref{fig:isr} together constitute clear evidence in support of the hypothesis that the increased suppression of large-radius jets with multiple skinny subjets as the separation $\Delta R_{12}$ between the hardest subjets in a plasma with $L_{\rm res}=\infty$ arises because with increasing 
$\Delta R_{12}$ it becomes increasingly likely that the subjets are increasingly likely to originate from different parton showers, with at least one originating from ISR.

\section{Imaging Large-Radius Jet Wakes in Inclusive Jet Events}
\label{app:inc}

In Section~\ref{sec:substructure}, we studied the shapes of the wakes produced by reclustered $R = 2.0$ jets with two skinny subjets selected by virtue of being produced in association with a high-$p_T$ $\gamma$, which we referred to as a semi-inclusive $\gamma$-jet ensemble. Clearly, the same analysis can 
just as well be carried out for an inclusive sample of jets, again $R=2.0$ jets composed of two skinny subjets. Since there will always be more inclusive jets than $\gamma$-tagged jets in any experimental data set, repeating our analysis for inclusive jets will benefit from higher statistics.  However, the interpretation of the observable that we have defined will be less crisp when it is applied to an inclusive sample of jets.  
In the $\gamma$-jet sample, in the hemisphere opposite to the two skinny subjets there will be the $\gamma$ that triggered the selection of the event, but it is unlikely that any jets will be found there.
In contrast, in an inclusive jet sample 
there will always be at least one other jet in the hemisphere opposite to the two skinny subjets.  And, the wake of this jet in the opposite hemisphere will yield a depletion of soft particles in the hemisphere where the two skinny jets ard found, in addition to an excess of soft particles in its own hemisphere. 
This depletion originating from the wake of the jet on the far side of the event will distort the wake of the two skinny subjets that is of interest to us in this paper, complicating the interpretation of the new jet shape observable that we have introduced. In this Appendix, we further explain, and illustrate, this complication.

Jets that traverse a droplet of QGP produced in a heavy ion collision will deposit momentum and energy into that droplet, exciting wakes that carry the deposited momentum and energy. Although some of the deposited energy goes into compression and rarefaction (i.e.~sound waves) most of the deposited momentum is carried by a region of the plasma behind the jet that has been pulled by the jet in its direction. Therefore, when one compares the freezeout of an unperturbed droplet of QGP to that of a droplet perturbed in this fashion, there will be a relative excess of soft particles in the direction of the boost experienced by the fluid cells, accompanied by a relative depletion of soft particles in the opposite direction. We call the excess of soft particles in the direction of the jet the ``positive wake" and the depletion of soft particles in the direction opposite the jet the ``negative wake". In the Hybrid Model, final state hadrons either originate from the fragmentation of the partons from the perturbative shower which have not lost all their energy after traversing the medium 
or are soft hadrons coming from the wake as described by Eq.~(\ref{eq:onebody}).
At the orientations where the latter expression is positive, it describes the soft hadrons that represent the positive wake; where it is negative,
it describes the soft hadrons that represent the negative wake~\cite{Casalderrey-Solana:2016jvj}.

Suppose that we have a heavy-ion collision event where we select a jet that is roughly back-to-back in the transverse plane with another jet, called an ``away-side jet". During jet reconstruction, the negative wake hadrons from the away-side jet will superpose with, and be clustered into, our selected jet (provided that they lie at similar rapidities~\cite{Pablos:2019ngg,Yang:2025dqu}). Since the jet shape observable in Eq.~\eqref{newjetshape} is linear in particle energy, we simply assign a negative energy weighting to the negative wake particles within a radius of $\Delta R = 2.0$ from the reclustered $R = 2.0$ jet's axis when calculating the jet shape. If no restrictions (selection criteria) are placed on the locations of away-side (sub)jets, then we will have no control over the locations of the negative wake hadrons from those away-side (sub)jets in $y$-$\phi$ phase space. In turn, we will have no control over where, in $y$-$\phi$ phase space, the negative wake hadrons are subtracted from the shape of our selected $R = 2.0$ jet in the hemisphere opposite the away-side (sub)jets. Therefore, the shape of the hadronic energy that is enhanced by two hard structures within in a reclustered $R = 2.0$ jet, will be affected by the shape of the hadronic energy that is depleted by (at least one) hard structures in the opposite hemisphere.

\begin{figure} [H]
    \begin{center}
    \subfloat[Wake shape for an ensemble of  inclusive jets, composed from only, and all, hadrons from jet wakes]{\includegraphics[width=1.0\textwidth]{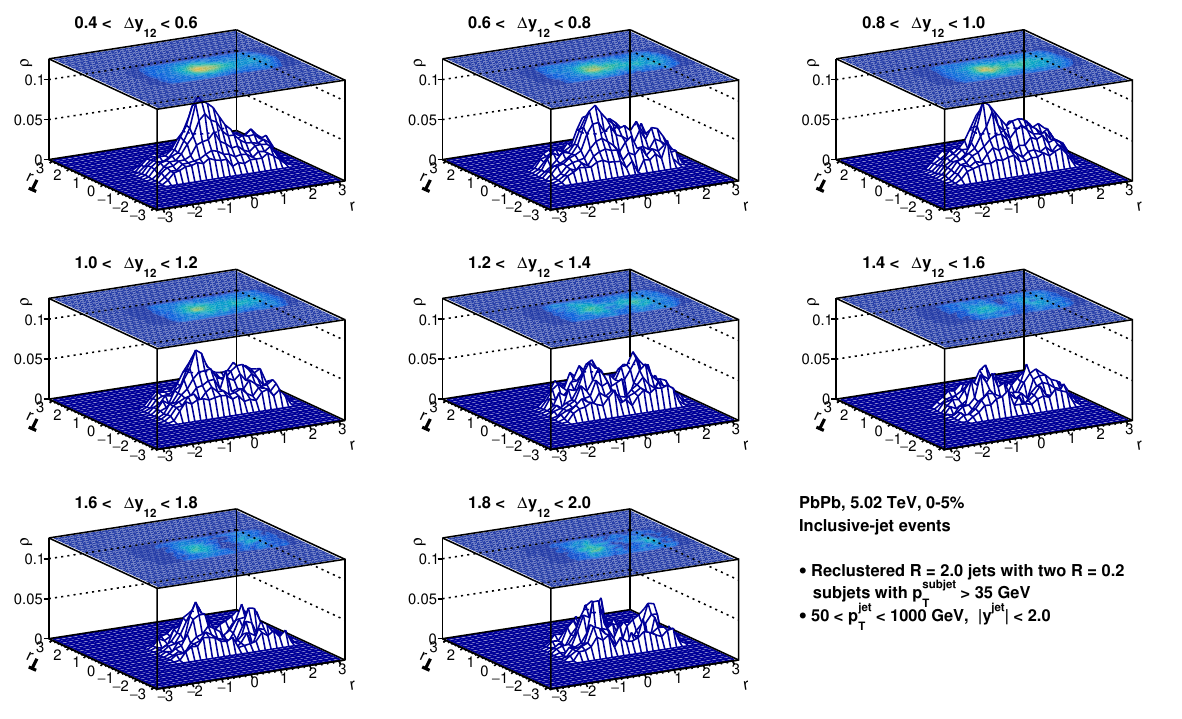}}\\
    \subfloat[Wake shape for the same jets as in (a), here including only the {\it positive} wake]{
    \includegraphics[width=1.0\textwidth]{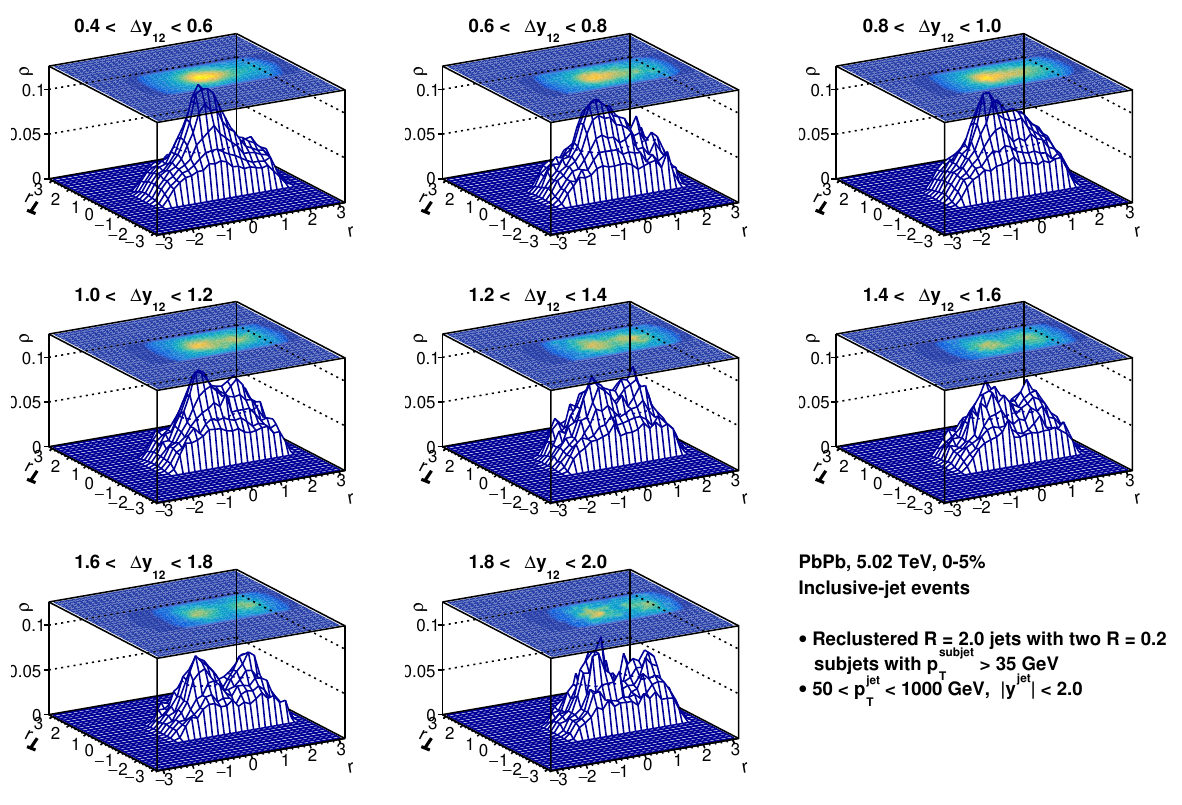}}
    \caption{(a) Jet shapes of reclustered $R = 2.0$ jets with two skinny anti-$k_t$ $R = 0.2$ subjets in PbPb collisions in inclusive-jet event, composed only from those hadrons originating from jet wakes within a radius of $\Delta R = 2.0$ around the reclustered $R = 2.0$ jet's axis. (b) Same as in (a) except that here the wake shape is composed only from those hadrons belonging to the \textit{positive wake}.}
    \label{fig:inclusive-wake-shapes}
    \end{center}
\end{figure}

We see the effects of this ``contamination'' in Fig.~\ref{fig:inclusive-wake-shapes}. Fig.~\ref{fig:inclusive-wake-shapes}(a) shows the 2-dimensional shapes of wakes (i.e. just the wake contribution) produced by reclustered $R = 2.0$ jets with two skinny subjets in inclusive jet events, with a $\widehat{p}_{T, \rm min}=15$ GeV. The jet-selection criteria used here is identical to the selection criteria used in our $\gamma$-jet analysis, but without any photon or away-side jet tagging. Furthermore, we have cut the distribution at $|r_{\perp}|<1$. The reason is that parts of the ``positive wake'' of the jets sitting at the opposite hemisphere enter the periphery of this large radius $R=2$ jet, severely impacting the $r$-projections. Note that the size of the wakes of these untagged jets sitting at the opposite hemisphere can be quite large, since we are not imposing any $p_T$ cuts on any possibly surviving jets momenta and thus one could be seeing fully hydrodynamized structures. By looking at Fig.~\ref{fig:inclusive-wake-shapes}(a), one observes the emergence of two peaks at around $\Delta y_{12}\gtrsim 1.0$, earlier than in the results shown in Fig.~\ref{fig:allparticleshapes}(b) for the $\gamma$-jet events.

The wake shapes in Fig.~\ref{fig:inclusive-wake-shapes}(a) are in fact diminished and distorted by the contribution from the negative wake(s) of jet(s) in the opposite hemisphere. We see this by comparing Fig.~\ref{fig:inclusive-wake-shapes}(a) to 
Fig.~\ref{fig:inclusive-wake-shapes}(b), in which we have plotted the jet shape observable composed only from hadrons originating from the positive wake. That is, in the Hybrid Model we can simply remove the confounding effects of the negative wakes of the jets in the opposite hemisphere.  Fig.~\ref{fig:inclusive-wake-shapes}(b) is the actual wake shape of the large radius jet with two skinny subjets that we wish to study; Fig.~\ref{fig:inclusive-wake-shapes}(a) is the contribution from all the wakes in the event, including the confounding effects from those of the jets on the opposite side.

\begin{figure} [t]
    \begin{center}
       {\includegraphics[width=1.05\textwidth]{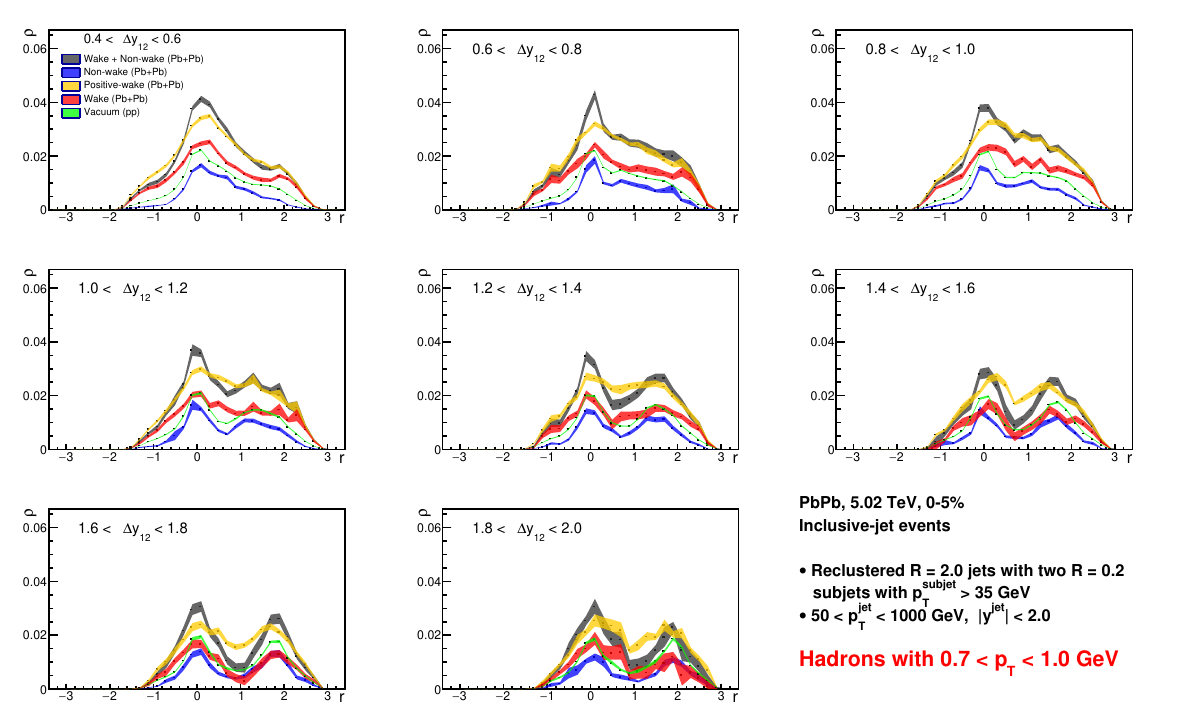}}
    \caption{Projections of the 2-dimensional jet shapes of large-radius $R = 2.0$ jets with two skinny subjets in inclusive-jet events onto the $r$-axis, composed only from hadrons with $0.7 < p_T < 1.0$ GeV within a radius of $\Delta R = 2.0$ around the large-radius $R = 2.0$ jet's axis.}
    \label{fig:inclusive-shapes-0.7-1.0}
    \end{center}
\end{figure}

To better understand these features and their implications for an experimentally realizable analysis of inclusive jets, we show in
Fig.~\ref{fig:inclusive-shapes-0.7-1.0} the projections of the full 2-dimensional jet shapes using only those hadrons with $0.7 < p_T < 1.0$ GeV onto the $r$-axis, calculated for collisions in vacuum (pp, in green) and PbPb collisions in inclusive jet events. As was done in Fig.~\ref{fig:1d-shapes-0.7-1.0}, the projections shown for PbPb collisions also differentiate between the shape of hadronic energy around the jet belonging to the wake (red) and the shape of hadronic energy around the jet coming from just the fragmentation of the jet (non-wake, in blue). There is an additional curve in this figure, namely the contribution of just the positive wake, in yellow. The effect of the negative wake originating from the untagged jets in the opposite hemisphere can be understood as the depletion observed when going from yellow to red (full wake). The positive wake comes solely from the wakes generated by the tagged subjets (since we cut away the positive wake of the untagged, recoiling subjets thanks to the $|r_{\perp}|<1$ cut), and represents the total of the wake distribution appearing in the plots in the main text for $\gamma$-jet samples (since the $\gamma$ does not produce wakes).

A first thing to note is that the sizes of the wakes are smaller for the inclusive jet samples studied in this Appendix than those for $\gamma$-jet samples in the main text. 
This arises because an inclusive jet selection involves selecting jets based on their own $p_T$,
and because of the steeply falling spectrum for jet production this necessarily biases the selection of jets towards those jets with a given $p_T$ that have lost the least energy.
In contrast, in the $\gamma$-jet sample the events are selected based upon the $p_T$ of the photon, which does not lose energy.  Selecting jets via their being found in events with a photon of a given $p_T$ does not bias the selection of jets toward those that have lost less energy.  This means that jets with a given $p_T$ in a $\gamma$-jet ensemble will on average have lost more energy than jets with the same $p_T$ in an inclusive sample, and so will have larger wakes.
For this reason, the relative contribution of the non-wake hadrons (blue) is larger in the inclusive jet samples in Fig.~\ref{fig:inclusive-shapes-0.7-1.0} than in the $\gamma$-jet analysis of Fig.~\ref{fig:1d-shapes-0.7-1.0}. This is a significant reason why studying $\gamma$-jet samples with this type of kinematic selections is especially appropriate 
when our goal is to visualize and analyze the 
properties of jet wakes~\cite{Brewer:2021hmh}.

On top of the fact that the wakes of the subjets of interest are smaller for the inclusive jets than in the $\gamma$-jet case, there is the depletion and distortion associated with the negative wake
originating from the jet(s) on the opposite side of the event, represented by the difference between the yellow and red curves in Fig.~\ref{fig:inclusive-shapes-0.7-1.0}.
We see that even though the positive wakes that we wish to study (yellow) present the appearance of two peaks at around $\Delta y_{12}\gtrsim 1.4$ in Fig.~\ref{fig:inclusive-shapes-0.7-1.0}, the total shape (gray) that would be measured in experimental data actually is more similar in shape to the non-wake (blue) distribution, 
because of the distortion introduced by the negative wake from the away-side jet in the 
total wake contribution (red) relative to the (yellow) wake of actual interest.  This 
distortion has so much of an effect that in the gray jet shapes there are already
two peaks at $\Delta y_{12}\gtrsim 1.0$.

In conclusion, due to both selection bias towards relatively unquenched jets, and the depletion effect caused by the negative wake produced by the strong quenching of recoiling hard objects,\footnote{This depletion effect is precisely what has been measured by CMS~\cite{CMS:2024fli} using $Z$-hadron correlations, 
which has been explained in terms of medium response physics such as hydrodynamic wakes.} inclusive jets do not represent the optimal scenario to visualize the shape of the wakes, even if one restricts to particles with low momenta. One possible way to mitigate the effect of the negative wake is by ``separating'' the recoiling jets, imposing a rapidity gap between the away-side jets and the large radius jet composed of skinny subjets that is of interest~\cite{Pablos:2019ngg}. This works because the wake stays fairly close in rapidity to the jet that produced it (unless one considers wakes which have had a lot of time to propagate before freezeout, in which case the distribution in rapidity can be broadened due to sound modes~\cite{Casalderrey-Solana:2020rsj}), and stems from the approximate longitudinal boost invariance of the medium and causality. While this (more statistics hungry) selection would make the total wake (red) signal as large as the positive wake (yellow) one, the modest relative size of the positive wakes (yellow) in the total shape would still disfavor employing inclusive jet samples when compared to what is possible with $\gamma$-jet samples, as shown in Section~\ref{sec:substructure}.



\section{Imaging the Two-Dimensional Shape of Large-Radius Jet Wakes with Hadrons in the Range $0.7<p_T<1.5$ GeV.}
\label{app:harderpt}

In Fig.~\ref{fig:1d-shapes-0.7-1.0}, we encountered the one-dimensional projections of reclustered $R = 2.0$ jet shapes onto the $r$-axis for collisions in vacuum (pp collisions) and PbPb collisions, calculated using only those hadrons with $0.7 < p_T < 1.0$ GeV. We saw that in this range of hadron-$p_T$, the wake (red) completely dominates the measured shape (gray) around the reclustered $R = 2.0$ jet's two skinny subjets. This makes imaging the wakes of such large-radius jets and imaging the substructure of these wakes in experiments a promising prospect.

\begin{figure} [H]
    \begin{center}
    \subfloat[Jet shape in PbPb collisions composed only from hadrons with $0.7 < p_T < 1.5$ GeV]{\includegraphics[width=1.0\textwidth]{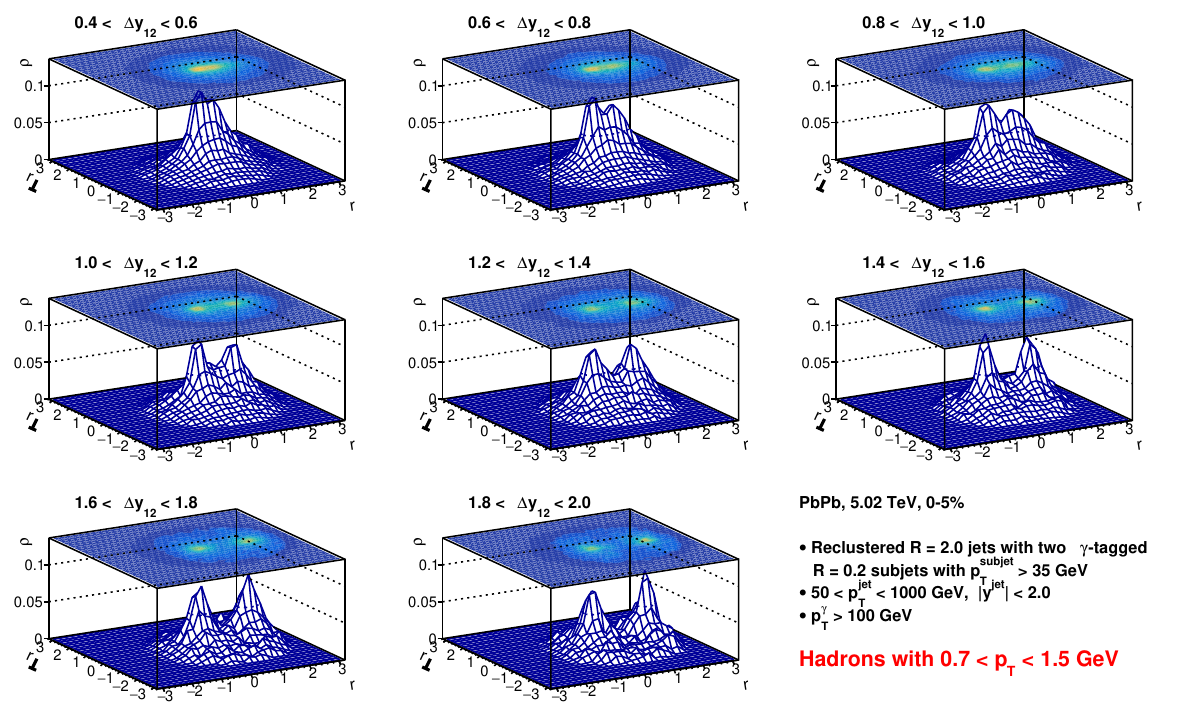}}\\
    \subfloat[$r$-projections of jet shapes composed only from hadrons with $0.7 < p_T < 1.5$ GeV]{\includegraphics[width=1.0\textwidth]{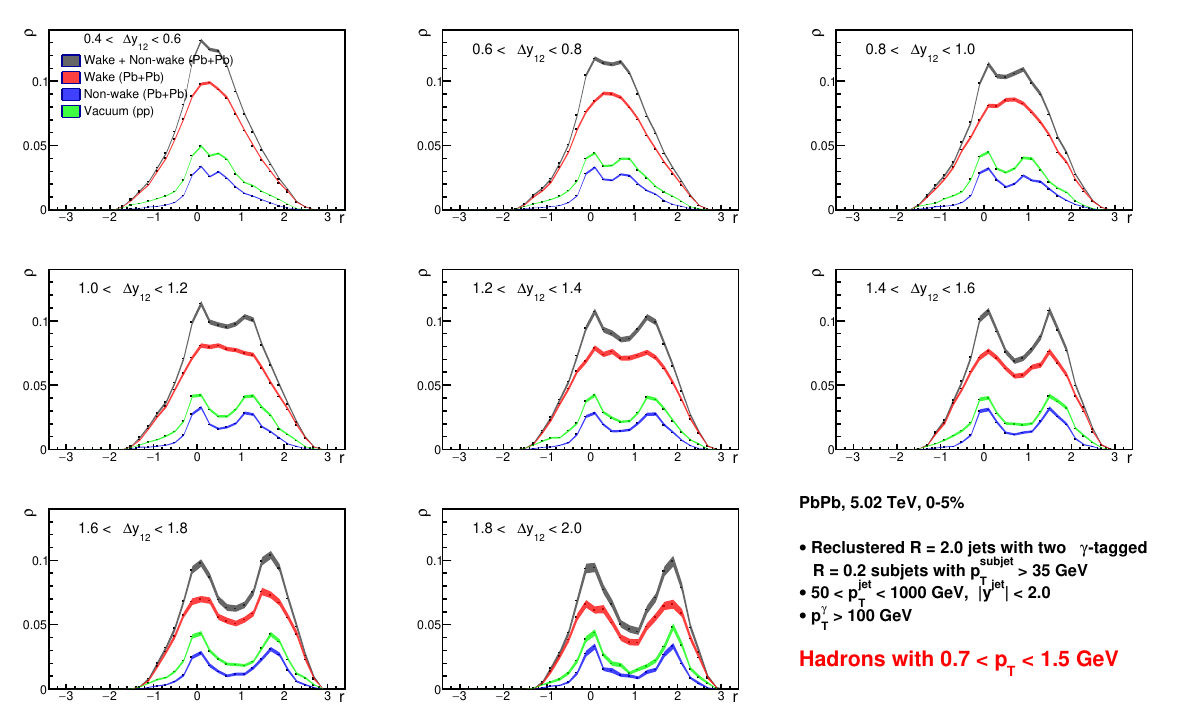}}
    \caption{(a) Jet shapes of reclustered $R = 2.0$ jets with two skinny anti-$k_t$ $R = 0.2$ subjets in PbPb collisions, composed only from those hadrons with $0.7 < p_T < 1.5$ GeV within a radius of $\Delta R = 2.0$ around the reclustered $R = 2.0$ jet's axis. (b) Projections of the 2-dimensional jet shapes of reclustered $R = 2.0$ jets with two subjets onto the $r$-axis, calculated using only those hadrons with $0.7 < p_T < 1.5$ GeV within a radius of $\Delta R = 2.0$ around the reclustered $R = 2.0$ jet's axis.}
    \label{fig:shapes-0.7-1.5}
    \end{center}
\end{figure}

In this Appendix we present results for the case in which we raise the upper cut on the momentum range of the hadrons entering the observable from 1 GeV to 1.5 GeV.
Fig.~\ref{fig:shapes-0.7-1.5}(a) shows the jet shapes of reclustered $R = 2.0$ jets containing two skinny subjets with different rapidity-separations $\Delta y_{12}$  in PbPb collisions, calculated using only those hadrons with $0.7 < p_T < 1.5$ GeV. Given the normalization used, this is actually equivalent to adding jet shapes calculated using hadrons with $1.0 < p_T < 1.5$ GeV to the jet shapes calculated using hadrons with $0.7 < p_T < 1.0$ GeV in Fig.~\ref{fig:2d-shapes-0.7-1.0}. Including the higher-$p_T$ hadrons 
results in the emergence of two distinct soft structures (for $\Delta y_{12} > 0.8$) before the wakes produced by the two subjets separate ($\Delta y_{12} > 1.2$) in Fig.~\ref{fig:allparticleshapes}. This is because the hadronic energy belonging to the wake represents a smaller relative fraction of the total hadronic yields in the range $1.0 < p_T < 1.5$ GeV. 
By comparing the results shown in Fig.~\ref{fig:shapes-0.7-1.5}(b), which show the jet shapes of reclustered $R = 2.0$ jets with two skinny subjets calculated using only hadrons with $0.7 < p_T < 1.5$ GeV and projected onto the $r$-axis, to the analogous results for the $0.7<p_T<1$ GeV range in Fig.~\ref{fig:1d-shapes-0.7-1.0}, one can observe that the relative contribution of non-wake hadrons (hadrons from fragmentation of the parton shower) is indeed increased, although keeping the lower cut of 0.7 GeV means that the wake still represents the leading contribution. 
Nevertheless, the emergence of the two peaks of the total shape at around $\Delta y_{12}\gtrsim 1.0$ in Fig.~\ref{fig:shapes-0.7-1.5}(b) cannot be attributed to the dynamics of the wakes, 
as the contribution from the peaked distributions of hadrons originating from the parton showers is too much of a confounding effect.


By comparing Fig.~\ref{fig:shapes-0.7-1.5}(a)
and Fig.~\ref{fig:1d-shapes-0.7-1.0}, both of which are experimentally realizable observables, to the wake shape in Fig.~\ref{fig:allparticleshapes}(b) composed only from all the hadrons originating from jet wakes (realizable in the Hybrid Model but impossible to realize experimentally) we see that Fig.~\ref{fig:1d-shapes-0.7-1.0}, 
composed from hadrons
with $0.7<p_T<1$ GeV, represents a good
proxy for the wake shape, much better than
Fig.~\ref{fig:shapes-0.7-1.5}(a) which
includes hadrons with $0.7<p_T<1.5$~GeV.
Fig.~\ref{fig:1d-shapes-0.7-1.0} represents a good
compromise that captures as much of the wake yields as possible, while diminishing the relative contribution of hadrons originating from fragmentation of the parton showers as much as possible. This allows for a direct visualization of the emergence of the two wakes in an experimentally realizable analysis. It would still be interesting to perform a scan of this observable in various $p_T$ ranges for a more complete characterization of this multiple subjet system, as already performed for single inclusive observables by ATLAS~\cite{ATLAS:2019pid} and dijet systems by CMS~\cite{CMS:2015hkr,CMS:2016cvr}.  


\bibliography{bibliography.bib}
\bibliographystyle{JHEP}

\end{document}